\documentclass{jfm}
\usepackage{graphicx}
\usepackage{amsmath}
\usepackage{amssymb}
\usepackage{natbib}
\usepackage{subcaption}
\usepackage{xcolor}

\newcommand{\rhobar}{\ensuremath{\overline{\rho}}}

\newcommand{\N}{\ensuremath{\mathit{N}}}
\newcommand{\Nbar}{\ensuremath{\overline{\N}}}
\newcommand{\bfez}{\ensuremath{{\bf e}_z}}
\newcommand{\bfv}{\ensuremath{{\bf v}}}

\newcommand{\eh}{\ensuremath{e_H}}
\newcommand{\Eh}{\ensuremath{E_H}}
\newcommand{\ev}{\ensuremath{e_V}}
\newcommand{\Ep}{\ensuremath{E_p}}
\newcommand{\Ea}{\ensuremath{E_a}}
\newcommand{\Eb}{\ensuremath{E_b}}
\newcommand{\avg}[1]{\ensuremath{\left<{#1}\right>}}
\newcommand{\avgh}[1]{\ensuremath{\left<{#1}\right>_H}}

\newcommand{\Fr}{\ensuremath{{Fr}}}

\newcommand{\Frbar}{\ensuremath{\overline{\mathrm{Fr}}}}

\newcommand{\Gn}{\ensuremath{{Gn}}}
\newcommand{\Rer}{\ensuremath{{Re_r}}}

\newcommand{\deltaU}{\ensuremath{\delta_{\cal U}}}
\newcommand{\deltaRho}{\ensuremath{\delta_{\rho}}}
\newcommand{\Sc}{\ensuremath{\mathrm{Sc}}}

\newcommand{\rhobard}{\ensuremath{\tilde{\rhobar}}}
\newcommand{\zd}{\ensuremath{\tilde{z}}}
\newcommand{\Ud}{\ensuremath{\widetilde{{\cal U}}}}
\newcommand{\rd}{\ensuremath{\tilde{r}}}

\newcommand{\elldH}{\ensuremath{\tilde{r}_m}}
\newcommand{\xd}{\ensuremath{\tilde{x}}}
\newcommand{\yd}{\ensuremath{\tilde{y}}}

\newcommand{\deltaUd}{\ensuremath{\tilde{\delta}_{\cal U}}}
\newcommand{\deltaRhod}{\ensuremath{\tilde{\delta}_{\rho}}}
\newcommand{\sd}{\ensuremath{\tilde{s}}}
\newcommand{\Ld}{\ensuremath{\tilde{L}}}

\newcommand{\nud}{\ensuremath{\tilde{\nu}}}
\newcommand{\Dd}{\ensuremath{\widetilde{{\cal D}}}}
\newcommand{\Nbard}{\ensuremath{\widetilde{\Nbar}}}
\newcommand{\gd}{\ensuremath{\tilde{g}}}

\newcommand{\eap}{\ensuremath{\epsilon_p}}

\newcommand{\epsilond}{\ensuremath{\widetilde{\epsilon}}}
\newcommand{\Nd}{\ensuremath{\widetilde{\N}}}

\newcommand{\atanArg}{\ensuremath{\frac{\varrho\rho}{\varrho^2-\rhobar\rho-\rhobar^2}}}
\newcommand{\Sv}{\ensuremath{S^2}}

\title[Wake in Non-Uniform Density Stratification]{Idealised Turbulent Wake With Steady, Non-Uniform Ambient Density Stratification}

\author[G. D. Portwood, S. M. de Bruyn Kops]
{G. D. Portwood$^{1,2}$, and S. M. de Bruyn Kops$^1$}

\affiliation{
  $^1$Department of Mechanical and Industrial Engineering,
  University of Massachusetts Amherst, Amherst, MA 01003-9284, USA \\
  $^2$ X-Computational Physics Division, Los Alamos National Laboratory, Los Alamos, NM 87545, USA
}

\date{\today}

\begin{document}
\maketitle

\begin{abstract}
  Density stratification in geophysical environments can be non-uniform in the vertical
direction, particularly in thermohaline staircases and atmospheric layer
transitions. Non-uniform stratification, however, is often approximated by the 
average ambient density change with
height. This approximation is frequently made in numerical simulations
because it greatly simplifies the calculations. In this paper, direct
numerical simulations using $4096 \times 2048 \times 2048$ grid points to
resolve an idealised turbulent wake in a non-uniformly
stratified fluid are analysed to understand the consequences of assuming
linear stratification. For flows with the same average change in density with
height, but varied local stratification $d\rhobar(z)/dz$, flow dynamics are
dependent on the ratio $\xi=\deltaU/\deltaRho$, where $\deltaU$ and
$\deltaRho$ are characteristic velocity and density vertical scale heights of
the mean flow.  Results suggest that a stably stratified flow will demonstrate
characteristics similar to nonstratified flow when $\xi > 2$, even though the
average stratification is quite strong.  In particular, the results show that
mixing is enhanced when $\xi > 2$.
\end{abstract}


\section{Introduction}
Geophysical flows can often be considered to have an ambient density that
varies with height and to be invariant on the time scales of turbulent motion.
Denoting by $\rhobar(z)$ the time-independent ambient density at vertical
position $z$, we define the ``stratification'' as $d\rhobar(z)/dz$.  Our
interest is in the stabilising stratification that occurs in much of the ocean
and atmosphere since understanding and modelling its effects on turbulence are
important to predicting weather and climate and the related turbulent
transport of heat, momentum and species.  In particular, we focus on the
dynamics of turbulence when the stratification is not uniform and on the
modelling biases that may occur if a non-uniformly stratified flow is modelled
as if it were uniformly stratified, that is, as if $d\rhobar(z)/dz$ were a
constant.

Numerous investigations have been performed concerning flow in density
stratified fluids. Laboratory studies include flows of turbulent wakes
\cite[e.g.,][]{cho93b,spedding96b,spedding02,bonnier02}, grid generated turbulence
\cite[e.g.,][]{liu95,fincham96,praud05}, and dynamics of monopoles and dipoles
\cite[e.g.,][]{billant00b,beckers01,beckers02}. Also, several numerical studies
have been performed to investigate how density-stratified fluids behave;
examples include horizontal layer decoupling
\citep{herring89,metais89,waite04}, turbulent mixing
\citep{winters96,staquet00,gargett03,peltier03}, turbulence parameterisation
\citep{smyth00a,shih05,hebert06a}, and flow energetics
\citep{ivey91,lindborg06a,almalkie12a,maffioli16,debk19,portwood19}.

\begin{figure}
\centerline{\includegraphics[width=10cm]{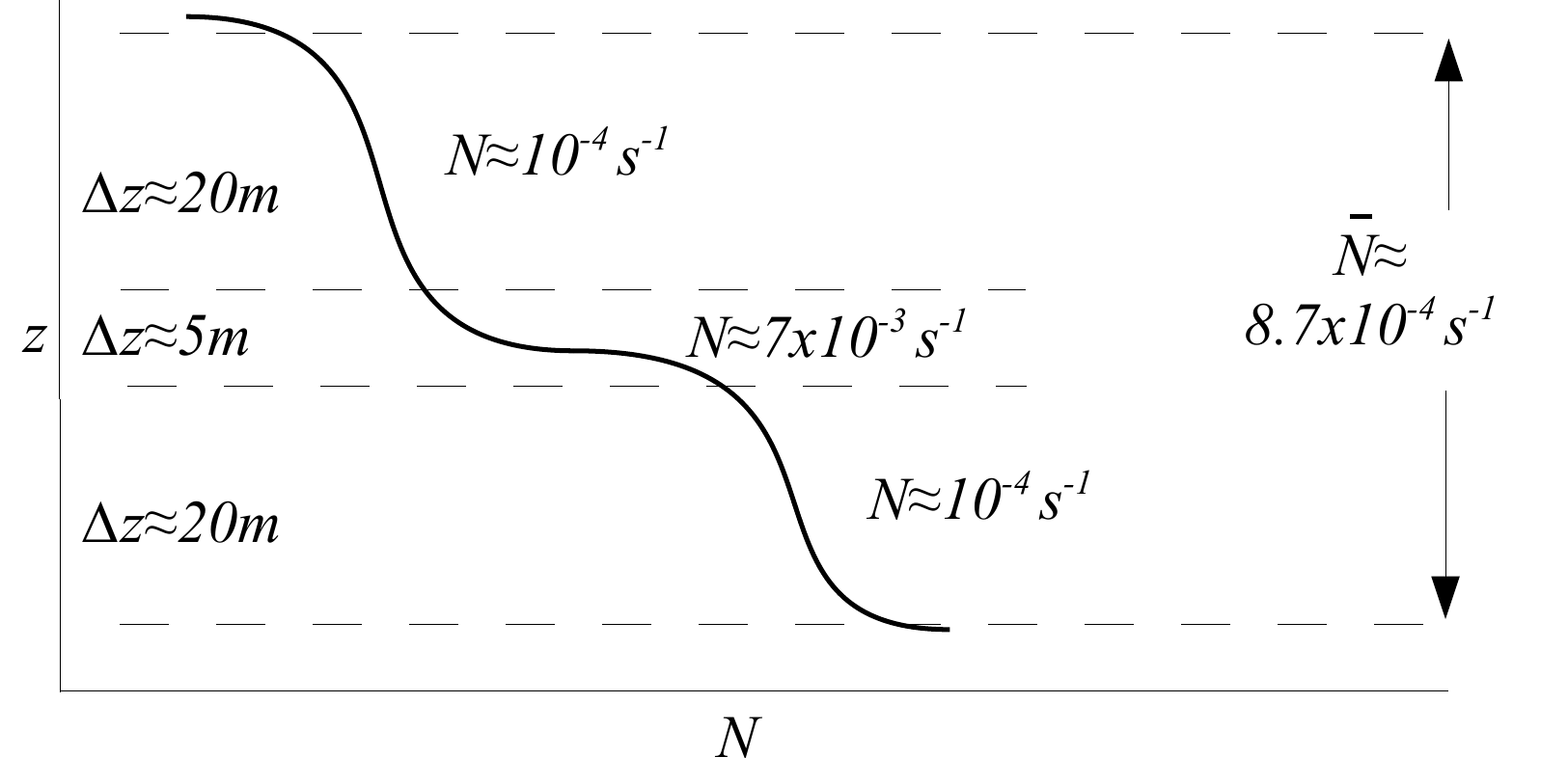}}
\caption{Cartoon of a density staircase showing two hypothetical layers subjected to
  moderate stratification separated by an interface of strongly stratified
  fluid.  The numerical values are from \citet{gregg87a}.
\label{fig:StairCartoon}}
\end{figure}

In the studies of density-stratified flows, such as those just cited, the
assumption that the stratification is uniform and not a function of height is
commonly made.  However, in natural settings such as the atmosphere
\citep{gossard85,dalaudier94,muschinski98} and the ocean
\citep{williams74,molcard77,lambert77,schmitt87,boyd89}, density layers or
``staircases'' can form in which areas of well-mixed fluid are separated by
thin, high density-gradient interface regions.  Assuming an average density
stratification $\Delta\rhobar / \Delta z = (\rhobar_{top}-\rhobar_{bottom}) /
(z_{top}-z_{bottom})$ in these regions neglects the effect of local
stratification differences.  In figure \ref{fig:StairCartoon} is a sketch
showing layers with values of the dimensional buoyancy frequency $N$ and its
spatial average $\overline{N}$ versus elevations $z$ typical of a thermohaline
staircase in the ocean in which the interface stratification can be 40 to 100
times larger than the layer stratification \citep{gregg87a}.

Previous studies have been performed to investigate shear flow in which the
initial vertical velocity and density profiles vary with height. For example,
\cite{smyth05} investigate the ratio of turbulent diffusivities,
\cite{smyth01} investigate a mixing efficiency defined as the
ratio of Ozmidov to Thorpe length scales, \cite{smyth00a,smyth00b} investigate
the length scales of turbulence in a shear flow, and \cite{dasaro04}
investigate mixing estimates using Lagrangian tracers.
In each of these
studies the initial density (and velocity) profiles are a function of vertical
position and change as the simulation evolves. 
An important difference between
those studies and the current research is that here the ambient vertical
density profile, $\rhobar(z)$, (and hence the density stratification) is
constant in time, thereby modelling a persistent, steady state thermohaline
staircase or atmospheric transition layer.

The objective of this study is to investigate the effect of non-uniform
density stratification on a highly idealised late wake.  In particular, the wake has
no source of energy so the flow is decaying. The wake has zero mean velocity,
similar to that generated by a self-propelled object (although
\cite{meunier06a} note that it is very difficult to obtain a truly
momentumless wake in a stratified fluid). The behaviour of the simulated flow
is investigated for a range of stratification profiles superimposed on the
same initial velocity field.  The methodology and simulations are described in
\S\ref{sec:methodology} followed by a detailed discussion of the dynamics of
the simulated flow in \S\ref{sec:general}, \ref{sec:location}, and \ref{sec:energetics}.
Modelling implications and conclusions are presented in
\S\ref{sec:conclusions}.
 
\section{Methodology}
\label{sec:methodology}
\subsection{Simulation Overview}
High resolution direct numerical simulations (DNS's) of a perturbed von
K\'{a}rm\'{a}n vortex street are conducted.  The initial conditions consist of
three vortex pairs and low-level noise.  There is no ambient shear and the the
ambient stratification is a function of height and held constant in time.
Each vortex is initialised with the following velocity profile
\citep[c.f.][]{debk03a}:
\begin{equation}  
{\bf \widetilde{V}_{\theta}} = {\cal
\Ud}\frac{\rd}{\elldH}\exp\left[\frac{1}{2}\left(1-
\left(\frac{\rd}{\elldH}\right)^2\right)\right]\mathrm{sech}^2\left(\frac{\zd}{\deltaUd}\right)
\ ,
\label{eq:utheta}
\end{equation}
where ${\cal \Ud}$ is the initial velocity scale, $\elldH$ is a radial length
scale, $\deltaUd$ is the velocity scale height, $\rd=\sqrt{\xd^2+\yd^2}$ is
the radial position, and $\zd$ is the vertical position.
The convention
of denoting dimensional quantities by $(\tilde{\cdot})$ is used in this section. 
The separation distances between vortex centres in the
$\xd$ and $\yd$ directions are $\sd_x=2\elldH$ and $\sd_y=1.5\elldH$. In
creating the initial flow condition, noise is applied to $\elldH$, $\deltaUd$,
$\sd_x$, $\sd_y$; each is randomly perturbed up to 5\% of its nominal
value. For example, the vertical scale for each vortex is calculated as
$\deltaUd + 0.05\lambda\deltaUd$, and the $y$ positions for the positive
vortices are calculated as $\Ld_y/2 + \sd_y/2 + 0.05\lambda \sd_y$, where
$\lambda$ is a [-1 1] uniformly distributed random number and $\Ld_y$ is the
span-wise ($y$) domain width.

The ambient density, $\rhobard(\zd)$, is imposed with a hyperbolic tangent
vertical profile,
\begin{equation}
\rhobard(\zd) = \frac{\Delta\rhobard}{2}
\mathrm{tanh}\left(\frac{-\zd}{\deltaRhod}\right) \ ,
\label{eq:rhobar}
\end{equation}
from which the density stratification, $d\rhobard(\zd)/d\zd$, is obtained:
\begin{equation}
\frac{d\rhobard(\zd)}{d\zd} =
-\frac{1}{2}\frac{\Delta\rhobard}{\deltaRhod}\mathrm{sech}^2\left(\frac{-\zd}{\deltaRhod}\right)\,.
\label{eq:drdz}
\end{equation}
Here $\Delta\rhobard = \rhobard_{top}-\rhobard_{bottom}$ is the difference in
density between the top and bottom of the numerical domain, $\zd$ is the
vertical position, and $\deltaRhod$ is a characteristic height of the density
profile.  The vertical profiles of both the velocity (\ref{eq:utheta}) and
density stratification (\ref{eq:drdz}) are $\mathrm{sech}^2$ with scale
heights $\deltaUd$ and $\deltaRhod$, respectively.  The parameter $\xi$ is
now defined which describes the ratio of the momentum vertical length scale,
$\deltaUd$, to the stratification vertical length scale, $\deltaRhod$, as
\begin{equation}
\xi\equiv\frac{\deltaUd}{\deltaRhod}.
\label{eq:xi}
\end{equation}
\begin{figure}
\centerline{\includegraphics[width=13cm]{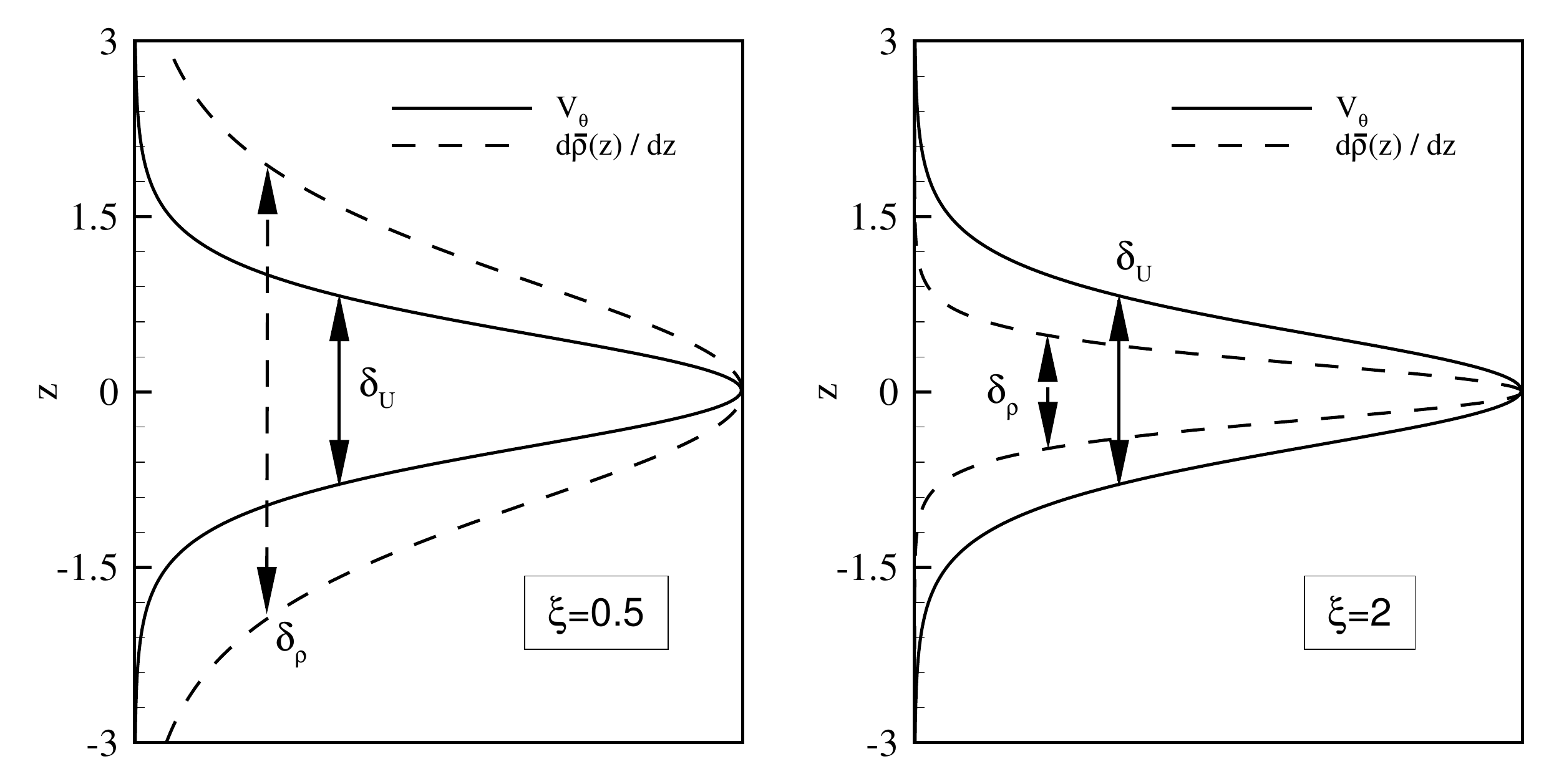}}
\caption{Velocity profile (solid) and density stratification (dashed) for
  $\xi=0.5$ and $\xi=2$.}
\label{fig:deltas}
\end{figure}
Velocity and density profiles corresponding to several values of $\xi$ are
shown in figure \ref{fig:deltas}.

\begin{figure}
\centering\includegraphics{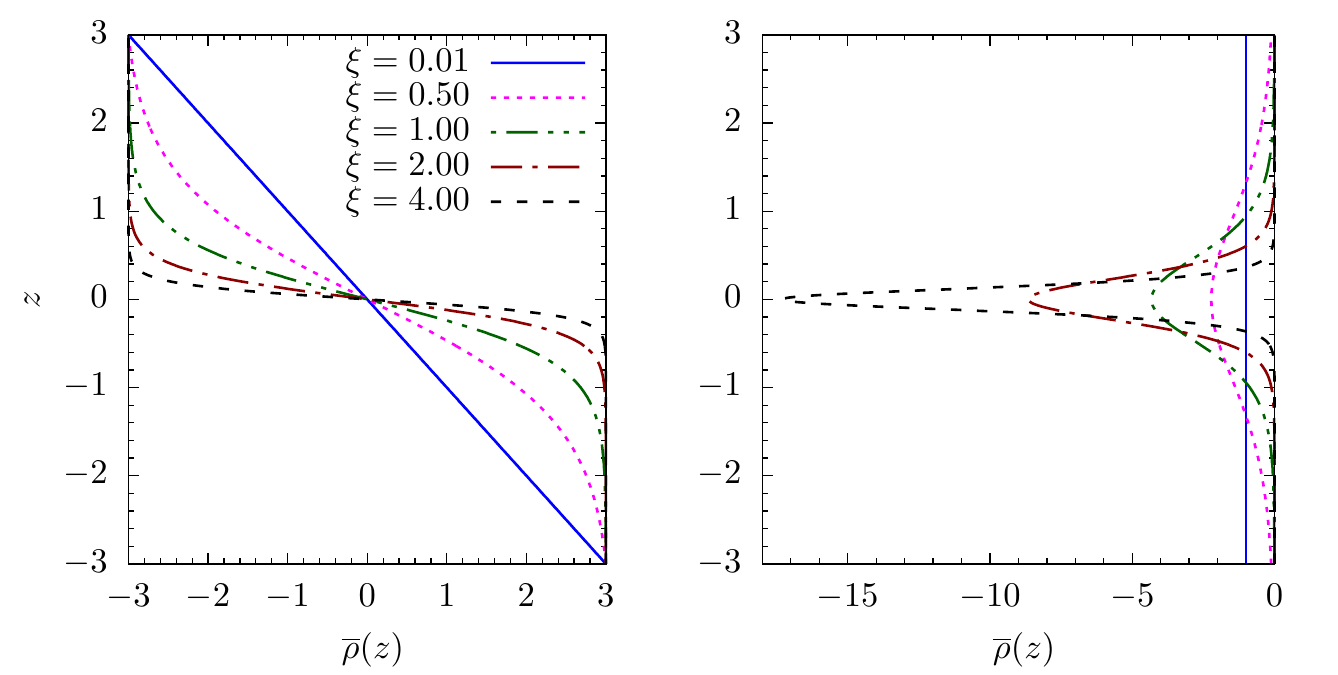}
\caption{Ambient (nondimensional) density profile and stratification
  profile for several different $\xi$ (see \eqref{eq:xi}). Note
  $\Delta \rhobar / \Delta z = -1$ for each $\xi$, but locally $d\rhobar(z) /
  dz$ is different.}
\label{fig:DRIC2}
\end{figure}

The velocity scale height is the same in all simulations; only $\deltaRhod$ is
varied to obtain different values of $\xi$.  The vertical profiles of
$\rhobar(z)$ and $d\rhobar(z)/dz$ for each $\xi$ are shown in figure
\ref{fig:DRIC2}. Note that while locally $d\rhobar(z)/dz$ differs for each
$\xi$, the average change in density with height is the same.

\subsection{Governing Equations}
\label{sec:goveq}
The simulated flow fields satisfy the Navier-Stokes equations subject to
the non-hydrostatic Boussinesq approximation. Taking $\Ud$ as the velocity
scale, $\elldH$ as the length scale, $\elldH|\Delta\rhobard / \Delta \zd|$ as
the density scale, $\elldH / \Ud$ as a time scale, and $\tilde{\rho}_0 \Ud^2$
as a pressure scale (where $\tilde{\rho}_0$ is the reference density value),
the nondimensional governing equations in a non-rotating frame of reference
are:
\begin{subequations}\label{eq:goveq}
\begin{equation}
\nabla \cdot \bfv = 0 \label{eq:cont}
\end{equation}
\begin{equation}
\frac{\partial \bfv}{\partial t} + \bfv \cdot \nabla \bfv = - \left(\frac{2
 \pi}{\Frbar} \right)^2 \rho \bfez - \nabla p + \frac{1}{\Rer} \nabla^2 \bfv
 \label{eq:NS}
\end{equation}
\begin{equation}
\frac{\partial \rho}{\partial t} + \bfv \cdot \nabla \rho +
w\frac{d\rhobar(z)}{dz} = \frac{1}{\Rer \Sc} \nabla^2 \rho \label{eq:density}
\end{equation}
\end{subequations}
where $\bfv = (u,v,w)$ is the velocity vector, $\rho$ and $p$ are the density
and pressure deviations from their ambient values, and $\bfez$ is a unit
vector in the vertical direction. The Reynolds, Froude, and Schmidt numbers
are defined as:
\begin{align}
\Rer &= \frac{\Ud \elldH}{\nud} & \Frbar&=\frac{2\pi \Ud}{\Nbard \elldH} & \Sc
&=\frac{\nud}{\Dd},
\label{eq:FRE}
\end{align}
where $\Nbard^2=-\gd/\tilde{\rho}_0(\Delta\rhobard / \Delta \zd)$ is the
average buoyancy  frequency, and ${\cal \Dd}$ is
an effective mass diffusivity that represents the effects of salt
diffusivity (ocean) or water vapour diffusivity (atmosphere) and thermal
diffusivity.

All simulations are conducted using a pseudo-spectral technique by which
spatial derivatives are computed using spectral methods. Time advancement is
performed using a third-order Adams-Bashforth scheme with pressure projection.
A spherical wave-number truncation of approximately 15/16 $\kappa_{max}$, with
$\kappa_{max}$ the maximum wave number in the discrete Fourier transforms, is
used to eliminate aliasing errors from wave numbers that are not already
affected by truncation error.  The extent of truncation error in high-resolution DNS
can be understood from the cusp in the spectra in, e.g., figure 5 in
\citet{kaneda03}, which is explained in \citet[\S5]{jang07}.
The momentum equation is
advanced in time with the nonlinear term computed in vorticity form. As
suggested by \citet{ker85}, an alternating time-step scheme is employed for
the density field to approximate the skew-symmetric form of the non-linear
term, thereby minimising aliasing errors \citep{boyd01}.

Periodic boundary conditions are imposed in all directions for the
simulations. This can be done because the stratification is very close to
periodic in the vertical direction.  To be precise, the stratification in the
simulations is not exactly $\mathrm{sech}^2$ but rather a Fourier series
approximation to it.  Since the governing equations include density
stratification $d\rhobard(\zd) / d\zd$, and not $\rhobard(\zd)$, periodic
boundary conditions can be used.

The only difference in the parameters of each simulation is the
stratification profile; $\delta_\rho$ is chosen so that $0.01 \le \xi \le 4$.  In
other words, the velocity vertical length scale ranges from 100 times smaller to 4
times greater than the density vertical length scale  (see figure
\ref{fig:DRIC2}).  Values of $\deltaRho$ and $\deltaU$ for each simulation are
summarised in Table \ref{table:sims}.  The case of $\xi = 0.01$ is so close to
being linearly stratified over the range of $z$ in the simulations that a
separate linearly stratified case is not considered.  The unstratified case is
also considered for comparison since, when $\xi \gg 1$, much of the flow is
subjected to very mild stratification.
\begin{table}
\begin{center}
\begin{tabular}{ccccccc}
$\xi$ & 0.01 & 0.5 & 1 & 2 & 4 & NoStrat\\
\hline
$\deltaRho$ & 69.4 & 1.39 & 0.694 & 0.347 & 0.174 & N/A\\
 $\deltaU$ & 0.694 & 0.694 & 0.694 & 0.694 & 0.694 & 0.694
\end{tabular}
\caption{Conditions for simulated flows}
\label{table:sims}
\end{center}
\end{table}

For all the simulations, $\Frbar=2.75$ and $\Rer=19200$.  Importantly, $\Sc$
is set to unity, close to the ratio of momentum to heat diffusivity in air but
far from the Schmidt number for salt in water. The nondimensional
computational domain size for each simulation is $L_x=12$ and $L_y=L_z=6$,
while the number of grid points in each direction is $N_x=4096$,
$N_y=N_z=2048$. This results in a ratio of the grid spacing to the average
Kolmogorov length scale of about six when the dissipation rate peaks in time,
which satisfies the resolution requirement of \citet{eswaran88}.  As discussed
in \citet{debk15}, when designing simulations of stratified turbulence there
is a tradeoff between resolving the dissipation range, resolving larger scales
strongly affected by buoyancy, and providing some scale separation between the
two.  In these simulations, as much of the resolution as practical is devoted
to providing that scale separation; the
large scales are represented by the minimum number of vortex pairs (three)
that can freely interact in a periodic domain.


\section{General Flow Features}
\label{sec:general}
\begin{figure}
  \centering
  \begin{subfigure}[b]{0.45\textwidth}
    \centering
    $t=0$
    \includegraphics[width=6.25cm]{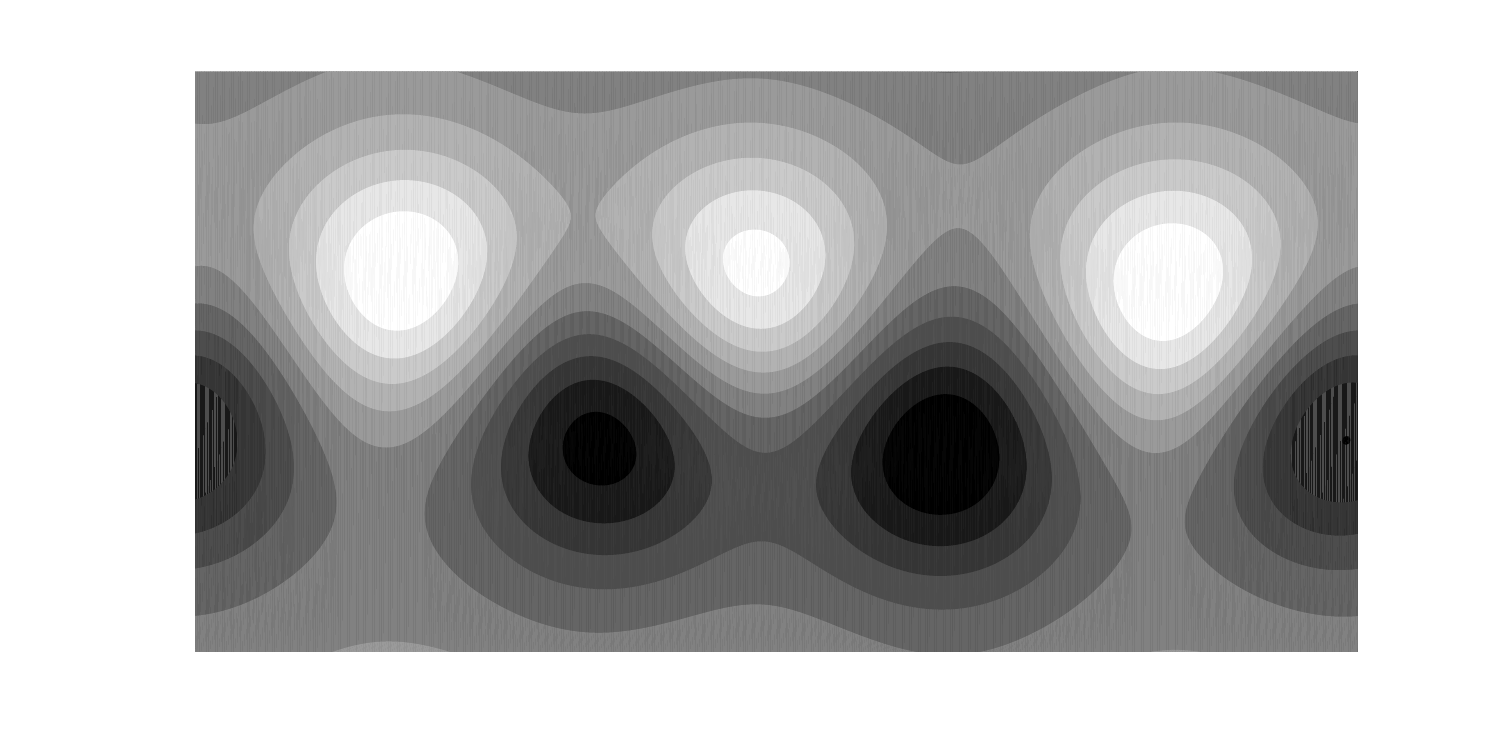}
  \end{subfigure}
  \hfill
  \begin{subfigure}[b]{0.45\textwidth}
    \centering
    $t=5$
\includegraphics[width=6.25cm]{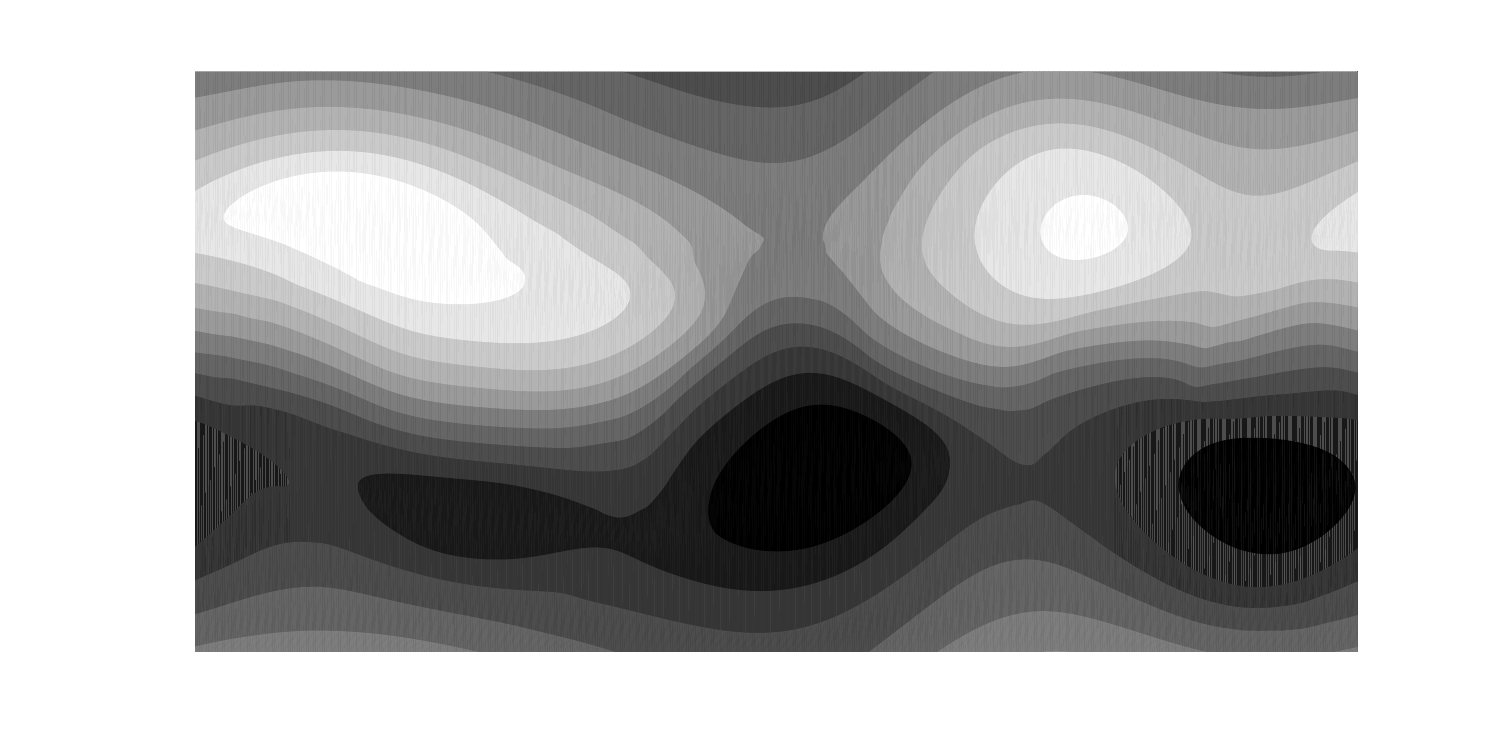}
  \end{subfigure}
  \vskip\baselineskip

  \begin{subfigure}[b]{0.45\textwidth}
    \centering
    $t=10$
\includegraphics[width=6.25cm]{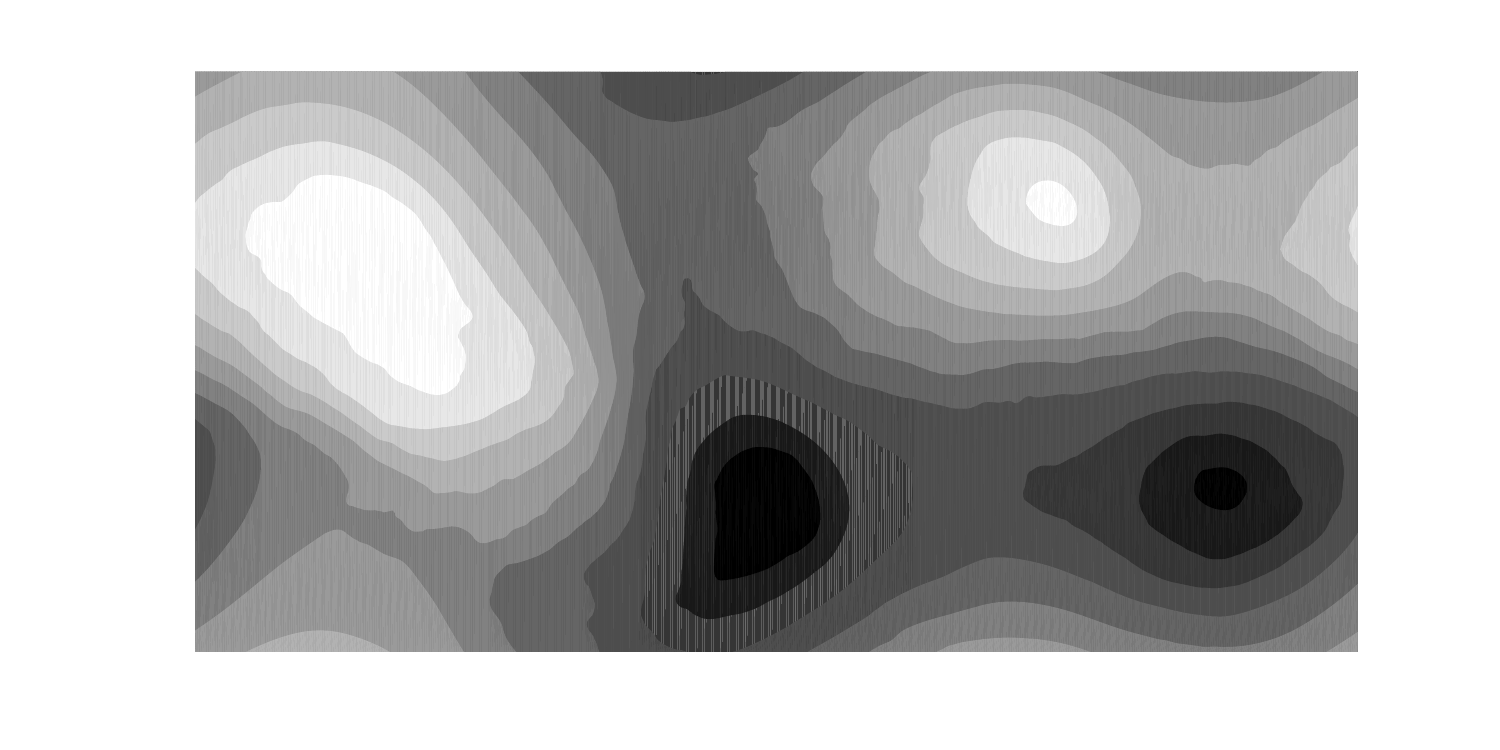}
  \end{subfigure}
  \hfill
  \begin{subfigure}[b]{0.45\textwidth}
    \centering
    $t=15$
\includegraphics[width=6.25cm]{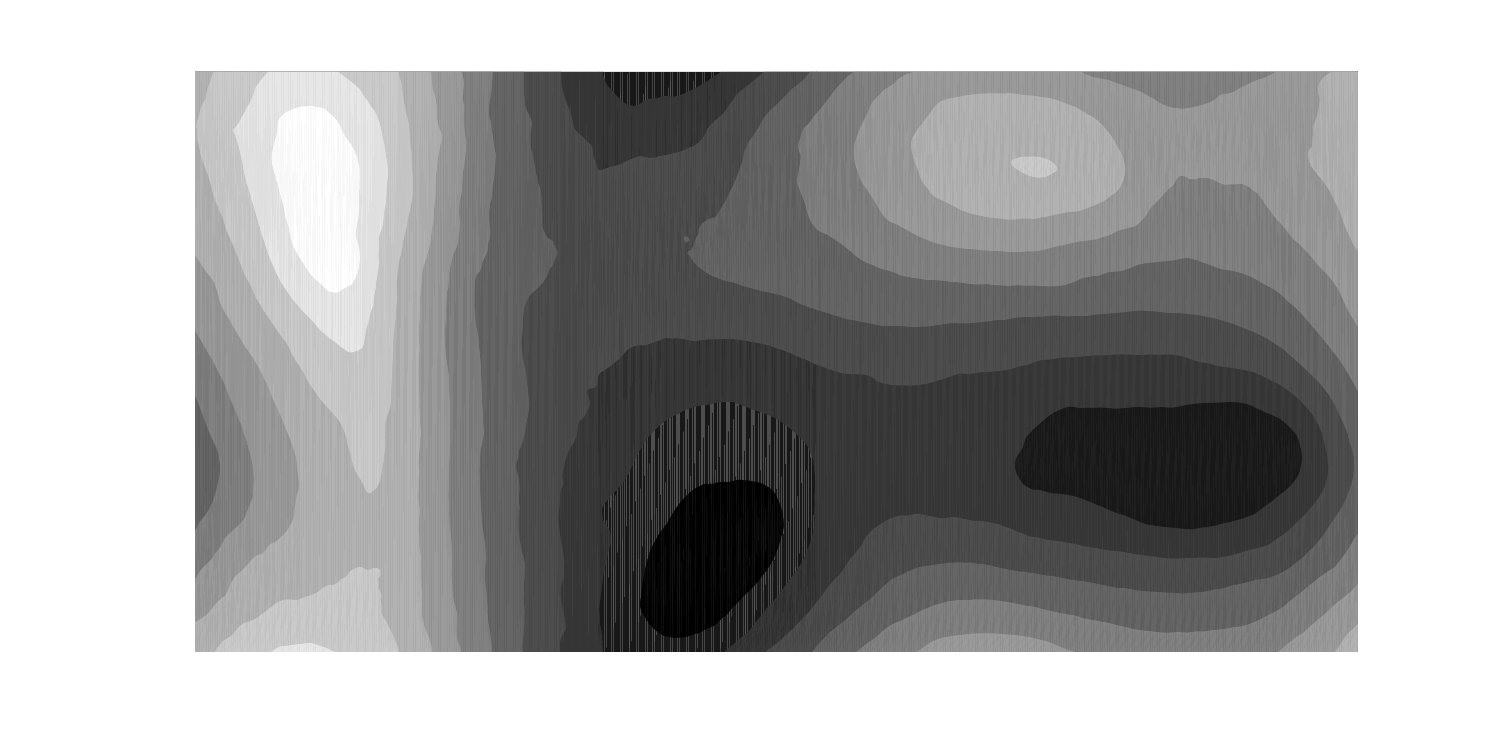}
  \end{subfigure}

\caption{Contour plot of centre plane stream function $\psi$ for
  $\xi$=0.01. Light colours represent positive values of $\psi$, dark 
  colours represent negative values.} 
\label{fig:uhevolDR100}
\end{figure}

Insight into the general flow dynamics can be obtained via use of the
horizontal stream function, $\psi$ \citep{riley00}
$$
\widehat{{\bf v}} = i(\bfez \times {\boldsymbol \kappa})\widehat{\psi},
$$ where ${\boldsymbol \kappa}$ is the three dimensional wave number and
$\widehat{(\cdot)}$ denotes a Fourier transformed quantity. A time-series plot
of the centreline $\psi$ is shown in figure \ref{fig:uhevolDR100} for the case
with $\xi$=0.01.  Cases having other $\xi$ values are qualitatively
similar. At $t$=0 the von K\'{a}rm\'{a}n vortex street is clear with
light colours representing positive vortices and dark colours representing
negative vortices. As the simulated flow advances in time the vortices
interact with each other and vortex pairing occurs by $t=15$.

\begin{figure}
\centerline{\includegraphics[width=10cm]{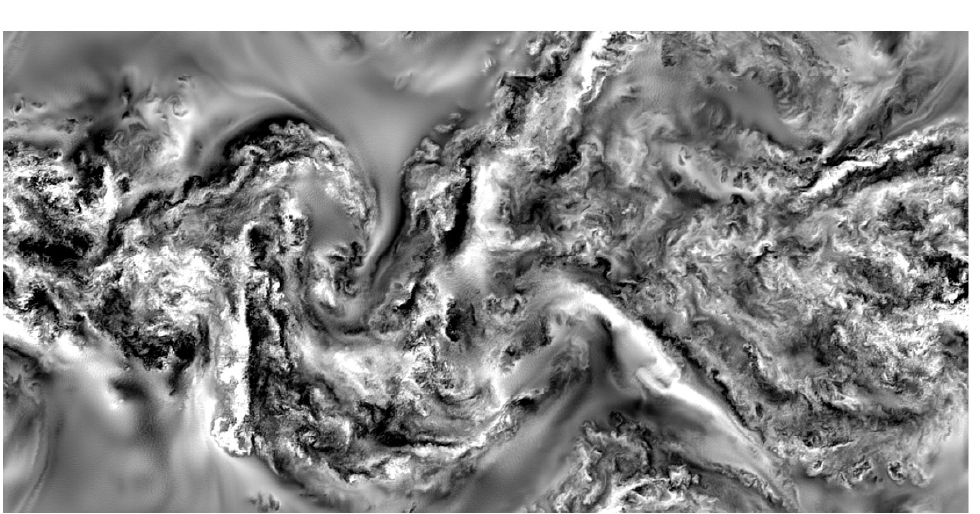}}
\centerline{\includegraphics[width=10cm]{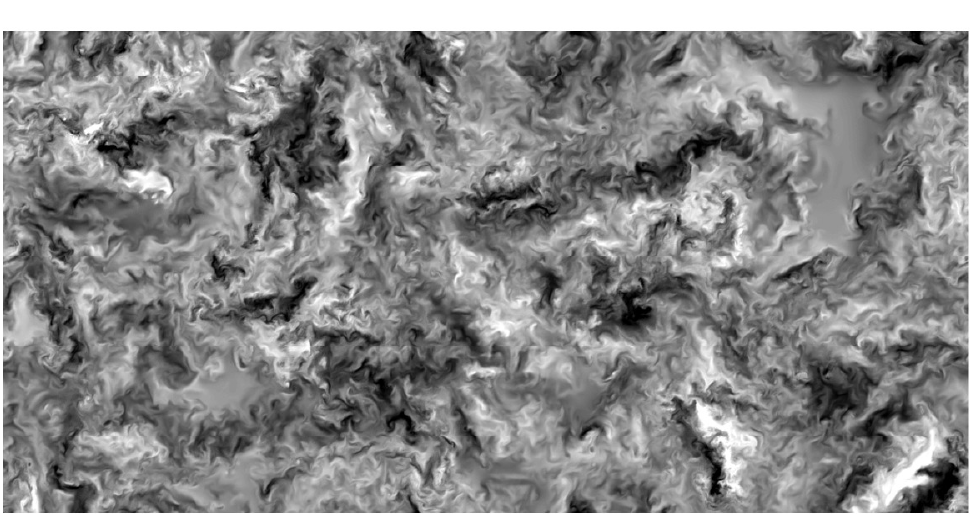}}
\caption{Centre plane ($z=0$)
  vertical velocity for $\xi=0.01$ (upper plot) and no stratification
  (lower plot) at $t=10$.  Dark colours represent negative (downward) velocity,
  light colours represent positive velocity.} 
\label{fig:u3evolDR100}
\end{figure}

As the flow evolves, vertical velocity $w$ can be generated by internal waves
and instabilities including shear \citep[e.g.,][]{holmboe62,lilly83, riley03}
and ``zig-zag'' \citep{billant00b, billant00a} instabilities. Since the flow
was initialised with $w=0$, it can be used as an indicator of regions of
turbulence generation. Figure \ref{fig:u3evolDR100} contains plots of vertical
velocity for $\xi=0.01$ (upper plot) and no stratification (lower plot) at
$t=10$. For the stratified flows, intermittent turbulent patches form
consistent with prior studies regarding density stratified flows.
\begin{figure}
  \includegraphics{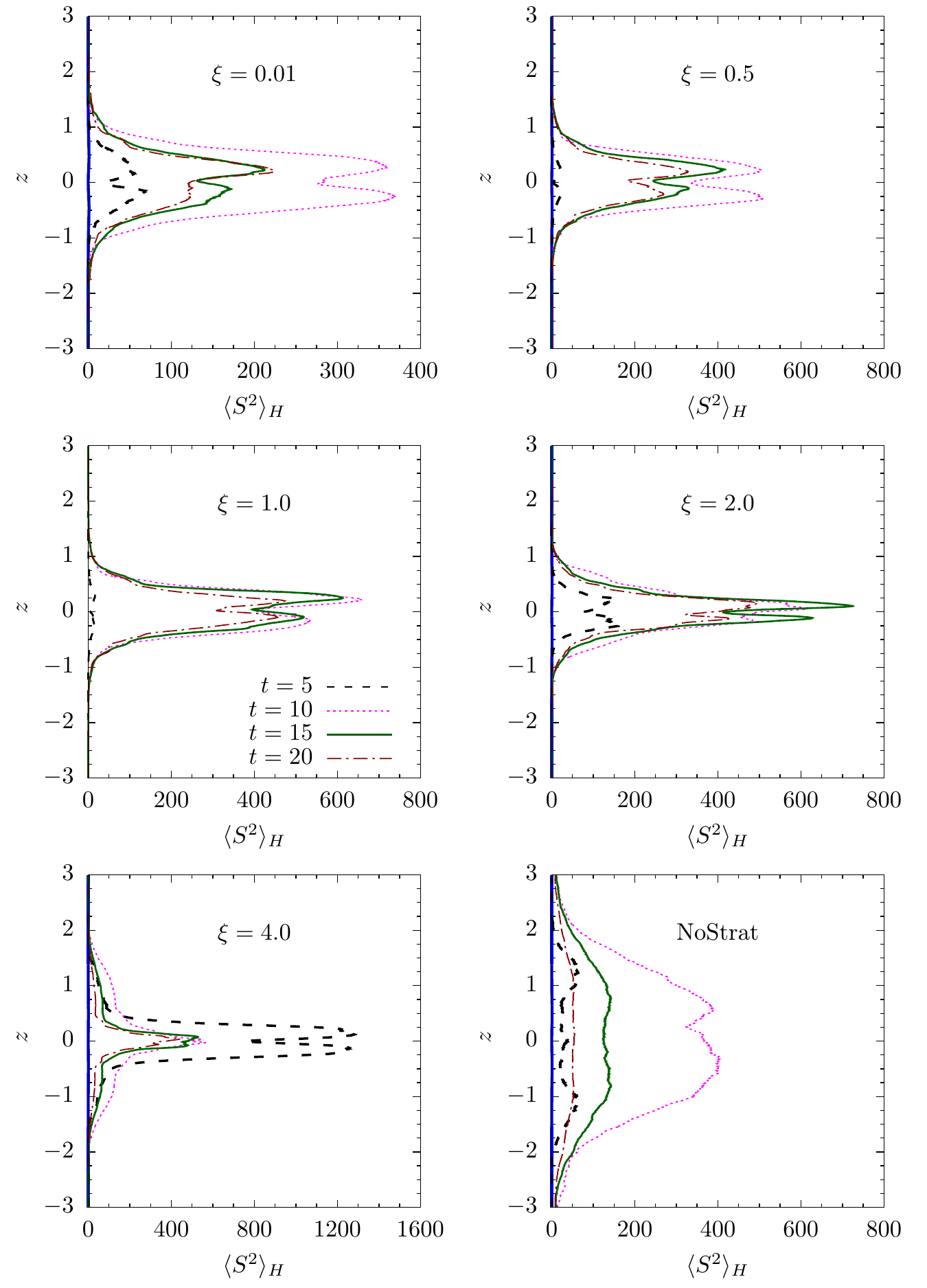}
\caption{Time evolution of $\avgh{\Sv}$ for several $\xi$ and no
  density stratification simulations. Note that the horizontal scale is not the same for all the plots.}
\label{fig:VShear}
\end{figure}
In stratified flows dominated by vortical modes, such as those simulated in
this study, it is postulated that horizontal layer decoupling occurs and that
the flow will be susceptible to Kelvin-Helmholtz shear instabilities
\citep{lilly83}. The square of the vertical shear of horizontal motions is
defined as:
\begin{equation}
\Sv = \left(\frac{\partial u}{\partial z}\right)^2 + \left(\frac{\partial
v}{\partial z}\right)^2\, .
\label{eq:VShear}
\end{equation}
Figure \ref{fig:VShear} contains plots of the horizontally averaged square of
vertical shear $\avgh{\Sv}$ versus vertical position for several $\xi$ and the
non-stratified simulation, where $\avgh{\cdot}$ denotes averaging over
$xy$-planes. Initially, the shear profile demonstrates a bimodal pattern,
with maximum values at $z \pm 0.5$, and is a result of the initial sech$^2(z)$
velocity profile.  For all the stratified cases the bimodal maximum shear
pattern persists in time, but the peaks do not remain aligned with the peaks
in the initial shear profile. Instead, the distance between peaks decreases in
time as the peaks move toward the planes of
maximum energy. This shift in the location of the maximum shear from that of
the maximum shear in the initial conditions toward that of the maximum energy
in the initial conditions is consisted with the results of \citet{riley03},
and suggests that the peak shear is due to decoupling of the horizontal
motions as suggested by \cite{lilly83} and not just from the initial
conditions.  In contrast, the bimodal pattern disappears by $t=5$ for the
non-stratified case in which horizontal decoupling does not occur.

\section{Location of Flows in Parameter Space}
\label{sec:location}
It has long been recongised that interpretation of simulation and laboratory
results depends on understanding the flow regime they are in.  For example, a
flow that is marginally turbulent may provide limited information about flows
at high Reynolds number.  In stratified turbulence, non-dimensionalisation of
the governing equations shows there are three dimensionless parameters which
we take to be a Reynolds, a Froude, and a Schmidt number in writing
\eqref{eq:goveq}.  Recently, \citet{debk19} showed that, for a given Schmidt
or Prandtl number, the Froude number and activity parameter, also called the
buoyancy Reynolds number, form a parameter space that is effective for
interpreting a large number of historical laboratory experiments and recent
simulations.  Understanding the location of the current simulations in this
parameter space is the subject of this section in order to prepare for
interpreting the flows in \S\ref{sec:energetics}

The motivation for choosing Froude number and activity parameter as the axes
on our parameter space for interpreting results is that turbulence is
characterised by a range of length scales.  In stratified turbulence, there
can be scales of motion strongly effected by buoyancy, strongly affected by
viscosity, and not affected significantly by either.  Let us consider ratios
of three length scales denoted $L$, $L_o$, and $L_k$. The most appropriate
definitions of these may be open questions, but conceptually they describe,
respectively, the scale of the turbulent kinetic energy, the largest scale
that is not directly affected by buoyancy, and the dissipation scale.  In this
context, an inverse Froude number is indicative of the scale separation
between $L$ and $L_o$ while the activity parameter is indicative of the scale
separation between $L_o$ and $L_k$.  Particularly in simulations, since the
total scale separation is quite limited, it is informative to consider length
scale ratios when evaluating simulation results.

For specificity, and for comparison with the literature, let us define
\begin{equation}
\tilde{L} = \frac{\tilde{k}^{3/2}}{\epsilond} \: , \hspace{1cm} \tilde{L}_o =
\left(\frac{\epsilond}{\Nd^3(\zd)} \right)^{1/2}\: ,  \hspace{1cm}
 \tilde{L}_k=\left(\frac{\tilde{\nu}^3}{\epsilond}\right)^{1/4} 
\end{equation}
where $\Nd^2(\zd) = \gd/\tilde{\rho}_0 d\rhobard(\zd)/d\zd$ is the buoyancy
frequency at vertical position $\zd$,
so that
\begin{equation}
\Fr = \left(\frac{\tilde{L}}{\tilde{L}_o}\right)^{-2/3} = \frac{\epsilond}{\Nd(\zd) \tilde{k}} \:, \hspace{1cm}
\Gn  = \left(\frac{\tilde{L}_o}{\tilde{L}_k}\right)^{4/3} = \frac{\epsilond}{\nud \Nd^2(\zd)} \ .
\label{eq:FrAndGn}
\end{equation}
$\Fr$ is the turbulence Froude number while $\Gn$ is the activity parameter,
also called the buoyancy Reynolds number.  In some literature, the
latter term is used for the product of the square of the horizontal Froude
number and the horizontal Reynolds number, introduced by \citet{riley03},
which is not the same as $\Gn$ unless certain inertial scaling assumptions
hold \citep[c.f.][]{debk19}.  To avoid confusion between the two quantities,
we use the original name, activity parameter, for $\Gn$
\citep[e.g.][]{dillon80}; the symbol is in deference to the introduction of
this quantity by \citet{gibson80} and to its identification by
\citet{gargett84} as a measure of the dynamic range available for turbulence
at scales too small to be directly affected by buoyancy.

The observation that stratified flow dynamics depend strongly on $\Fr$ and
$\Gn$ has been the subject of many studies.  For example $\Fr$ must be
sufficiently small for the scaling arguments of \citet{billant01} and
\citet{riley03} to hold.  \citet{brethouwer07} describe a `strongly'
stratified regime corresponding to $Fr \sim O(0.01)$ or smaller, and \citet{falder16}
refer to this regime as `layered anisotropic stratified turbulence' (LAST) to avoid
ambiguity of the meaning of `strongly stratified.'
\citet{diamessis11} report strong effects of buoyancy on stratified
turbulent wakes for values of $\Fr$ (inferred from their data) significantly
above that of the LAST regime.  \citet{almalkie12a} report the
characteristics of homogeneous stratified turbulence in the LAST regime and
less strongly stratified, and provide a conversion chart between various
Froude numbers for their data.  Note that the conversion between $L$ and the
integral, or correlation, length scale is dependent on Froude number
\citep{debk15,maffioli16} so that a generic conversion between Froude numbers
is not practical.

The history of the importance of $\Gn$ to parametrise the dynamics of
stratified turbulence goes back farther.  Using scaling arguments,
\citet{gibson80} estimated that $\Gn \approx 24$ is the minimum value for
``active'' turbulence to form.  This estimate was soon followed with ocean
measurements showing that markedly differing flow regimes correspond to
different ranges of $\Gn$ \citep{gargett84}.  Since then, multiple studies
have been focused on the importance of correctly accounting for the effect of
$\Gn$, and not just for Froude and Reynolds number, in simulations
\citep{smyth00b,almalkie12a,bartello13}.  Based on simulations with
domain-averaged $\Gn=$ 13, 50, and 220, \citet{portwood16} propose modeling a
stratified turbulent flow as an amalgamation of flow regimes distinguished by
the $\Gn$ averaged over subregions of the flow.
They conclude that regions with locally averaged $\Gn$ of $O(1)$, $O(10)$, and
$O(100)$ are dynamically distinct, which is consistent with the ocean
measurements of \citet{gargett84}.  Recently, \citet{debk19} use new
simulations and historical laboratory data to show that the time-evolution of
decaying flows depends strongly on $\Gn$.  In particular, viscous effects are important
when $\Gn \sim 10$ and dominate when $\Gn \sim 1$.  Finally, we note that the
stratified late wakes of \citet{watanabe16} with $\Gn$ on centreline of
$O(10)$ evolve differently from simulated and laboratory wakes with lower
centreline $\Gn$.

\begin{figure}
\begin{center}
  \includegraphics{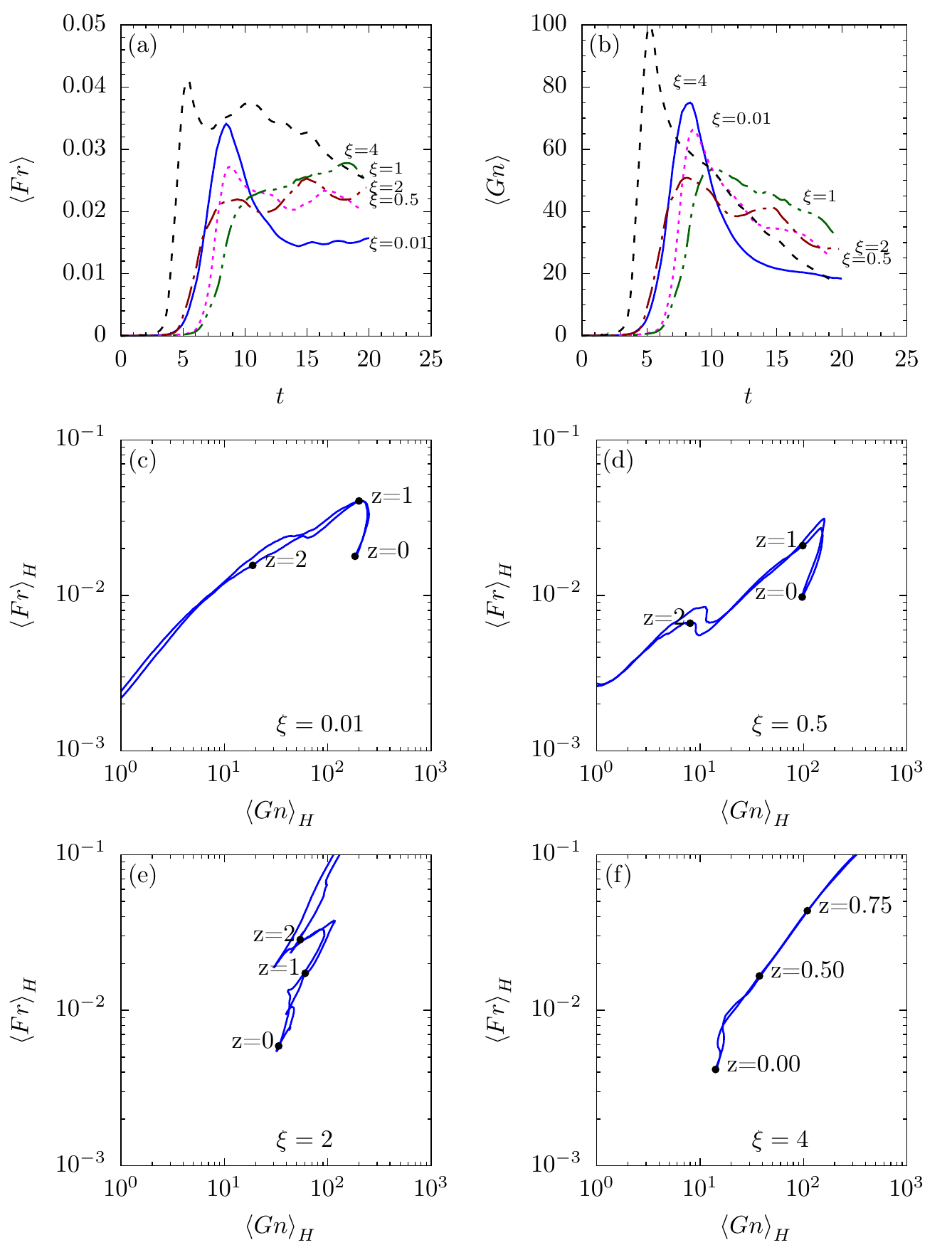}
\end{center}
\caption{Panels (a) and (b): time evolution of domain-averaged $\Fr$ and
  $\Gn$. Panels
  (c), (d), (e), (f): location of each plane in $\left<Gn\right>_H -
  \left<Fr\right>_H$ space at $t=10$ for four of the five cases with three planes
  marked with dots for each case; one curve in each panel is for the top half of the wake ($z\ge0$) and the other curve for the bottom half of the wake.
\label{fig:FrAndGn}}
\end{figure}

To understand $\Fr$ and $\Gn$ in the context of non-uniform stratification, we
consider domain-averages and planar-averages:
\begin{align}
\avg{Fr} &=  \frac{\avg{\epsilond}}{\Nbard \avg{\tilde{k}}} \ ,&
\avg{Gn}  &= \frac{\avg{\epsilond}}{\nud \Nbard^2} \ ,
\\
\avgh{Fr} &=  \frac{\avgh{\epsilond}}{\tilde{N} \avgh{\tilde{k}}}\ ,  &
\avgh{Gn}  &= \frac{\avgh{\epsilond}}{\nud \Nd^2(\zd)} \ .
\label{eq:AvgFrAndGn}
\end{align}
  The time evolutions of the domain
averaged quantities are plotted in figure \ref{fig:FrAndGn}.
The plot of $\left<Fr\right>$ in figure
\ref{fig:FrAndGn}(a) might suggest that the flows are strongly stratified but perhaps
not in the LAST regime, and to the plot of $\left<Gn\right>$ that the flows
consist of patches of turbulence by the criteria of \citet{portwood16}.  These
conclusions are called into question, though, after reconsideration of how $N$ and
$\epsilon$ vary in the vertical. For this reason, parameter maps for four of
the cases at $t=10$ are included in figure \ref{fig:FrAndGn} as panels
(c), (d), (e), and (f).  Time $t=10$ is chosen because there has been sufficient
time for the flows to develop but they are still highly energetic
(c.f.\ figure \ref{fig:u3evolDR100}).  From these panels, it is apparent that
there are regions of the flow in the LAST regime ($Fr \lesssim O(0.01)$) and
regions not in the LAST regime both over a significant range of $Gn$.

Figure 7 gives our first understand of fundamental difference between flows
with nearly uniform stratification and those with sharply varying
stratifications.  In panels (c) and (d) of the figure, which are the flows
with the most nearly uniform stratification, $Gn$ is highest on the centreline
and decreases toward the edges.  $Fr$ is also highest in the wake core and
decreases toward the edges.  This is because the stratification is significant
at the edges so that it, combined with the lack of wake energy, results in
very low turbulence.  In panels (e) and (f), which are flows with strong
variation in stratification, the opposite is true because for
$\xi=2$ and $4$ the stratification is so weak at the edges that the turbulence
is high.  In all cases, though, the location in the $\left<Fr\right>_H -
\left<Gn\right>_H$ plane indicate that all the wake cores are in the LAST regime
by the criteria of \citet{brethouwer07} and dominated by three-dimensional
turbulence by the criteria of \citet{debk15} and \citet{portwood16}.

\section{Energetics and Mixing}
\label{sec:energetics}
In figure 7 we observe that the layout of the wake in parameter space when $\xi$ is low
is different from that when $\xi$ is high.  This suggests that the mixing
characteristics will be different for different $\xi$.  To explore this, we
consider energy and mixing in this section.

\subsection{Kinetic Energy}
\label{sec:KE}
\begin{figure}
  \includegraphics{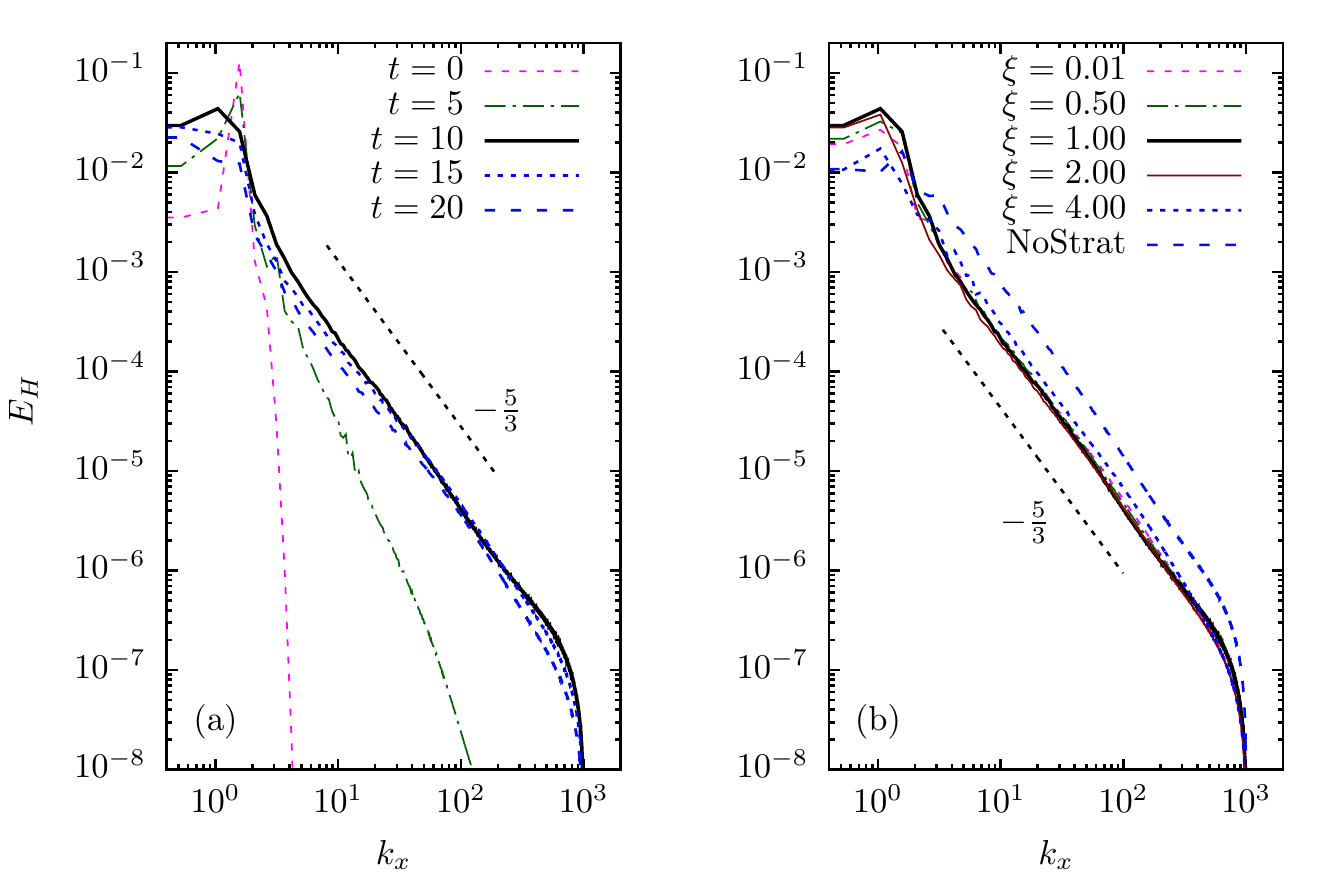}
\caption{(a) Evolution of $\Eh(k_x)$ $\xi=1$; (b) $\Eh(k_x)$
  $x$ spectrum for all $\xi$ at $t=10$. The curves in (b) for the stratified
  cases are almost indistinguishable except near the left edge where they go
  in this order from bottom to top: NoStrat, 4.00, 0.01, 0.50 1.00, 2.00.
\label{fig:ExSpec}}
\end{figure}

The horizontal and vertical contributions to kinetic energy are defined as:
\begin{equation*}
\eh = \frac{1}{2}\left(u^2 + v^2\right) \ , \quad \ev =
\frac{1}{2}\left(w^2\right) \ .
\end{equation*}
These are local quantities.
An overview of the flow energetics is given by the 
the $x$-direction spectrum of horizontal energy 
defined as
$$
E_H(k_x)=
\frac{1}{2}\left(\widehat{u}(k_x)\widehat{u}(k_x)^* +
\widehat{v}(k_x)\widehat{v}(k_x)^*\right)\, . 
$$ Here $\widehat{u}(k_x)$ and $\widehat{v}(k_x)$ are the $x$ and $y$
components of velocity Fourier transformed in the $x$-direction, $(\cdot)^*$
denotes a complex conjugate, and $k_x$ is the $x$-direction wave number.
Figure \ref{fig:ExSpec}(a) contains a plot of the time evolution of
$\widehat{E}_H(k_x)$ for $\xi=1$, which is representative of the spectra of
all simulations conducted in this study.  The initial peak at $k_x \approx
\pi/2$ corresponds to the separation distance between vortices.  As the flow
evolves, transfer of energy to smaller scales is observed as the magnitude of
$\widehat{E}_H(k_x)$ at larger wave numbers increases between $t=0$ and
$t=10$. After $t=10$, the flow is fully developed in the sense that the energy
at all wave numbers decreases for $t>10$.  In addition, once the flow has had
time to develop from the initial conditions ($t>10$), it displays a
$k_x^{-5/3}$ spectrum spanning a decade of $k_x$.  This $k_x^{-5/3}$
dependence is seen for all $\xi$ in figure \ref{fig:ExSpec}(b). Other
simulations have exhibited approximately $k_x^{-5/3}$ spectra
\citep{riley03,lindborg06a,bartello13}, but it has been observed that slopes
tend to be flatter than $k_x^{-5/3}$ for low Froude number
\citep{bartello13,debk15} although the results of \citet{debk15} and
\citet{debk19} indicate that low $Gn$ affects the spectral slope and that, in
sets of simulations, it is often difficult to distinguish the effects of $Fr$
from those of $Gn$.

\begin{figure}
  \includegraphics{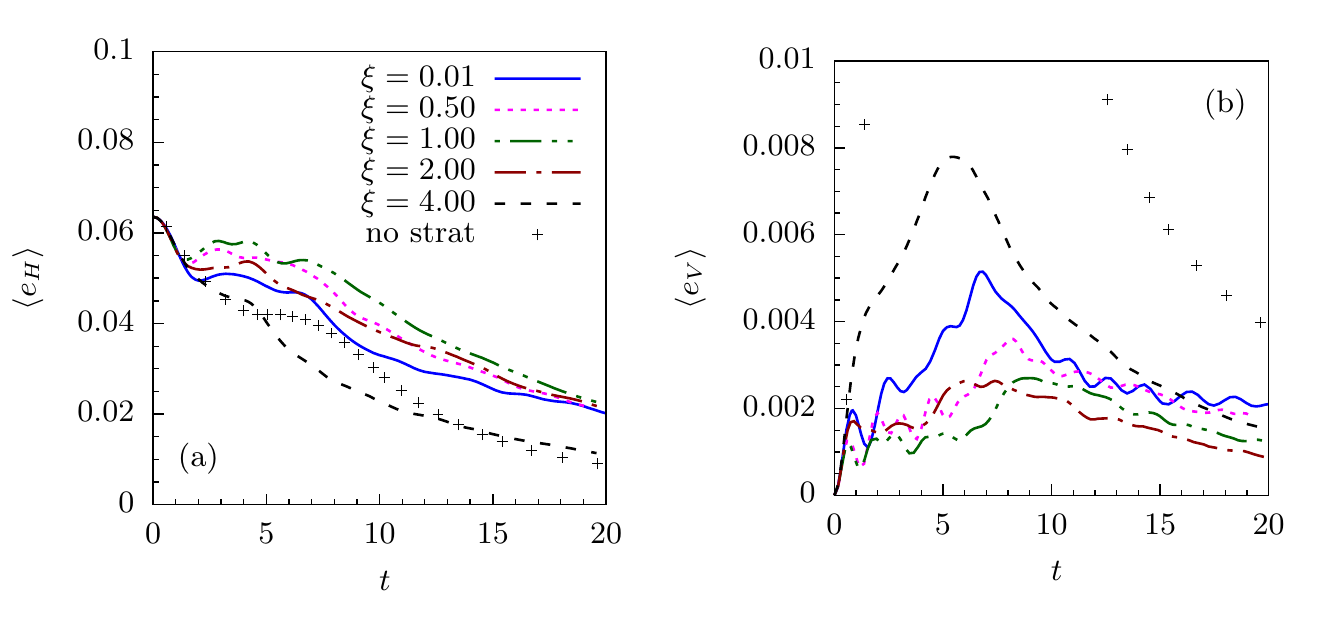}
\caption{Time evolution of the average (a) horizontal and (b) vertical contributions to kinetic energy.}
\label{fig:KEStrat}
\end{figure}

The time evolution of $\avg{\eh}$ and
$\avg{\ev}$ for each $\xi$ are shown in figure
\ref{fig:KEStrat}.
As $\xi$ is increased from 0.01 to 2, the trend is for $\avg{\eh}$ to persist
longer in time and the magnitude of $\avg{\ev}$ to decrease compared with the
unstratified flow.  In contrast, the case with $\xi=4$ loses very nearly the
same amount of energy between $t=0$ and $t=20$ as the unstratified cases, and
it loses approximately 83\% of its energy in this time compared with
approximately 58\% for the other cases.  The case with $\xi=4$ also creates a
lot of vertical motion as indicated by the plot of $\avg{\ev}$ in figure
\ref{fig:KEStrat}(b).

\begin{figure}
  \includegraphics{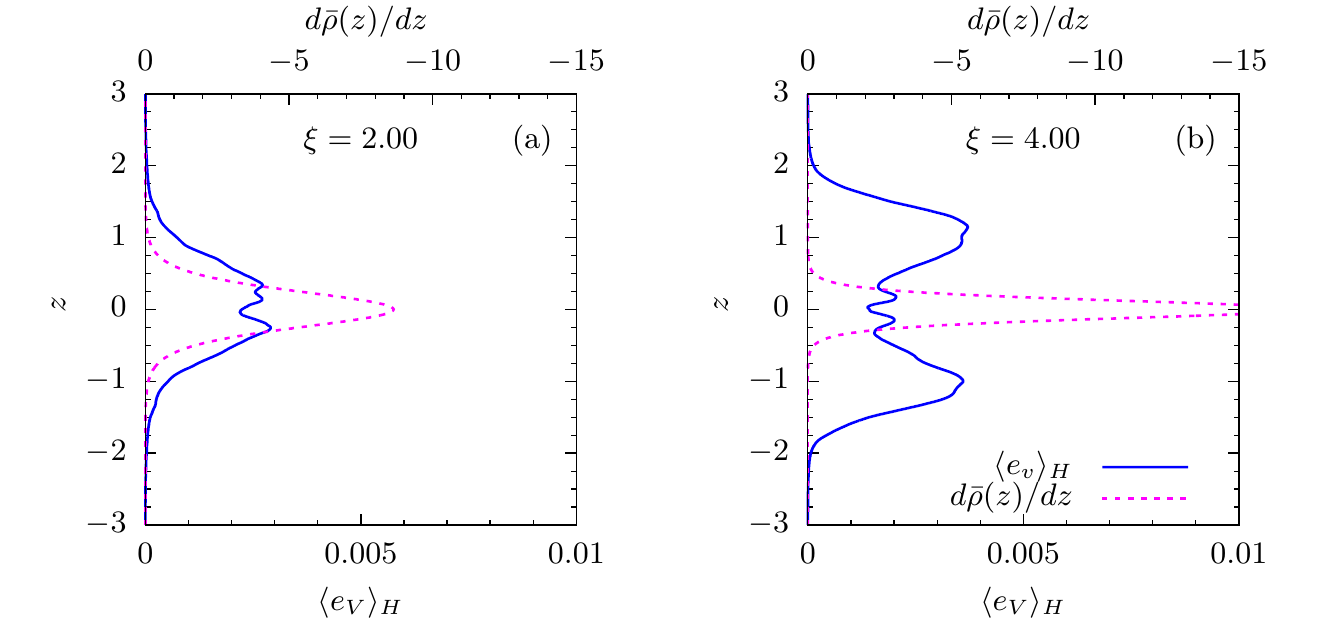}
  \caption{$\avgh{\ev}$ at $t=10$ for $\xi=2$ and $\xi=4$.
\label{fig:evStrat}}
\end{figure}

To understand why the case with $\xi=4$ evolves markedly differently from the
other cases, we present the vertical profiles of $\avgh{\ev}$ for $\xi=2$ and
$\xi=4$ in figure \ref{fig:evStrat}. With $\xi=2$, most of the vertical motion is
in the region where the mean density gradient is fairly strong whereas in
$\xi=4$ the vertical motion is strongest outside the strongly stratified
region.  It appears that with $\xi=4$ that the regions above and below the
strongly stratified zone act as unstratified wakes that decay faster than the
stratified wakes.

\subsection{Available Potential Energy}
\label{sec:APE}
Discussion of potential energy usually involves the concepts of available and
background potential energy introduced by \citet{lorenz55}. He noted that in
order to convert the total potential energy in the Earth's atmosphere to
kinetic energy, the temperature needed to reach absolute zero and all mass
needed to be located at sea level.  Instead, the potential energy that is
available for conversion to kinetic energy, $\Ea(t)$, is defined to be the result of
any deviation from a background (or reference) potential energy, $\Eb(t)$,
defined as the potential energy that
would occur if the fluid were adiabatically redistributed to a minimum energy
state. The available potential energy is the total potential energy, $\Ep(t)$,
minus the background potential energy:
\begin{equation}
\Ea(t) = \Ep(t) - \Eb(t) \, .
\end{equation}

$E_a$ is a time varying quantity for a volume of fluid and each fluid
element in the volume contributes to it and so one can define the local
available potential energy, $e_a$, such that
\begin{displaymath}
\Ea(t) = \int_{\mathcal V} e_a(\vec{x},t) d {\mathcal V}
\end{displaymath}
is the available potential energy for some volume ${\mathcal V}$.  Several
papers discuss how $e_a$ must be defined so that it is fully consistent with
the concept of available potential energy
\citep{holliday81,roullet09,molemaker10,winters13}.

A subtle concept is that $e_a$ is a local quantity that is
affected by non-local changes in the density field via the effect of those
non-local changes on the instantaneous reference state. 
Consider a fluid parcel with density 
$\rho_t(\vec{x},t)$.  If all the fluid parcels in some volume are adiabatically
sorted to find the reference configuration having minimum potential energy, $\rho_\ast(z,t)$,
then that fluid parcel will move to a new location $z_*$.  Alternatively, the
parcel is elevated a distance $\zeta = z - z_*$ from the location
it would occupy in the minimum energy arrangement.  The local available
potential energy is not proportional to $\rho \zeta$, though, because all the fluid
parcels with sorted locations between $z$ and $z_*$ affect $e_a(\vec{x},t)$.
Therefore,
\begin{equation}
e_a(\vec{x},t) = \left(\frac{2\pi}{\Frbar}\right)^2\left[ \zeta\rho(\vec{x},t) - \int_{z_*}^z
  \rho(z'_*) dz' \right] \ .
\label{eq:ea}
\end{equation}
This expression is equivalent to that derived by \citet{roullet09}, and
\citet{winters13} shows that it is positive semi-definite and integrates to
$E_a$.  We evaluate it numerically by sorting the density field with an
algorithm that allows $\zeta$ to be recovered and by computing the integral in
the expression via the trapezoid rule.

Before moving on to analysing potential energy in the simulations, let us define for
convenience some notation regarding densities. The total density is
\begin{equation}
\rho_t(\vec{x},t) = \overline{\rho}(z) + \rho(\vec{x},t)
\end{equation}
where an arbitrary additive constant has been omitted.
The excess density relative to the reference density is
\begin{equation}
\rho_e(\vec{x},t) = \rho_t(\vec{x},t) - \rho_\ast(z,t).
\end{equation}
If the flow is inviscid then $\rho_e = \rho$ and if the mean density profile is
given by \eqref{eq:drdz} then
\begin{equation}
e_a=\left(\frac{2\pi}{\Frbar}\right)^{2}
  \left[\rho_t\mathrm{arctanh}\left(\atanArg\right) +
  \frac{\varrho}{2}\ln\left(\frac{\varrho^2-(\rho_t)^2}{\varrho^2-\rhobar^2}\right)\right]
  \, ,
\end{equation}
where $\varrho = \Delta \rho / 2$, independent of time \citep{hebert07}.

\begin{figure}
  \includegraphics{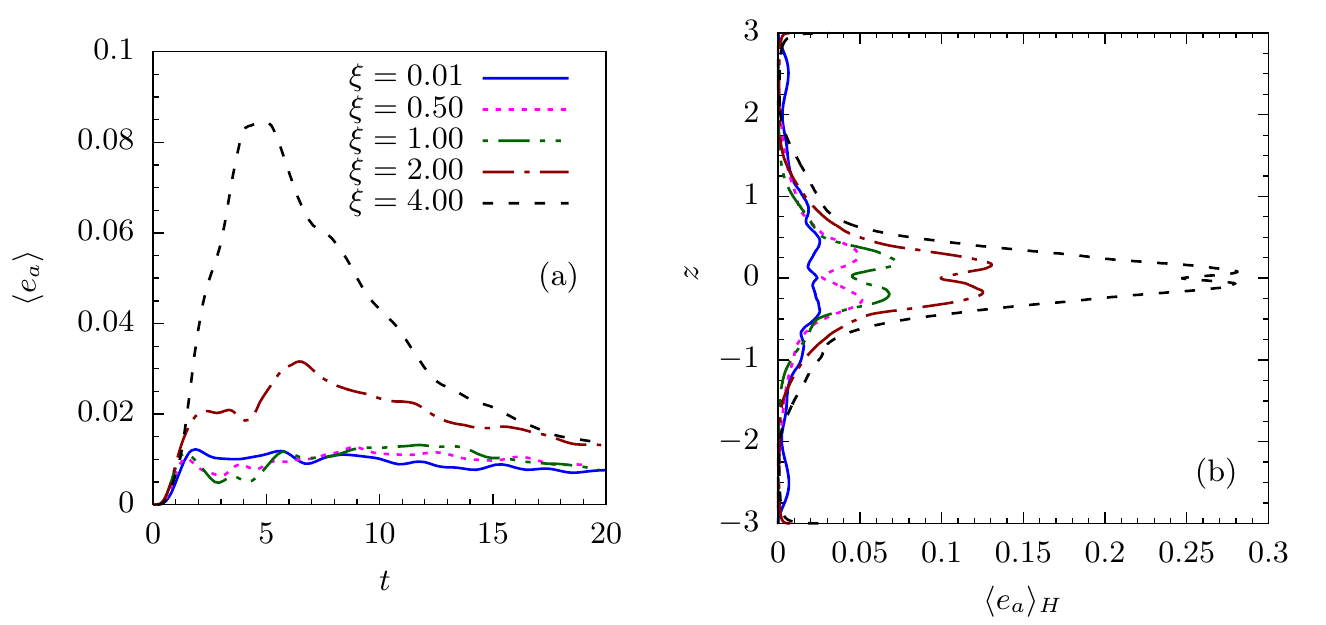}
\caption{Evolution of (a)  $\avg{e_a}$; (b) $\avgh{e_a}$ at $t=10$ for
   each $\xi$.  
\label{fig:epStrat}}
\end{figure}

The time evolution of $\avg{e_a}$ for each $\xi$ is shown in in figure
\ref{fig:epStrat}(a).
These data are computed by sorting the fields to
determine the background potential energy and then subtracting it from the
total potential energy.  The notation $\avg{e_a}$ is used in order to be
consistent with that for kinetic energy, but note that $\avg{e_a} = E_a(t)$.
Evident from the figure is that the case with $\xi=4$ generates significantly
more available potential energy than the other cases, which is the result of
the higher vertical motion in this case evident from figure \ref{fig:KEStrat}.

The vertical profiles of $\avgh{e_a}$ are shown in figure
\ref{fig:epStrat}(b).  The bulk of the available potential energy in all cases
is in the wake core.  This is interesting because we see in figure
\ref{fig:KEStrat}(b) that the majority of the vertical motion in the case with
$\xi=4$ is outside the strongly stratified core.  From figure
\ref{fig:epStrat}(b), though, we observe that since the stratification is weak
in the region of the strongest vertical motion there is little potential
energy associate with it.  We conclude that, even though most of the potential
energy is in the strongly stratified regions, the higher vertical motion with
$\xi=4$ results in higher potential energy.

Also from figure \ref{fig:epStrat}(b) we observe a weakness in our analysis
technique based on computing the local available potential energy $e_a$.  In
the paragraph preceding \eqref{eq:ea} it is observed that, while $e_a$ is a
local quantity, it is affected by all the other fluid parcels in the sorting
volume used for determining the background potential energy.  In this case, we
have defined the sorting volume as the entire computational domain, which is
entirely arbitrary because we could simulate the wakes in a
different size domain.  An artifact of this arbitrariness is that extreme tails in the
profiles of $\avgh{e_a}$ exhibit values that are higher than at some locations
closer to the center of the wake.  This happens because the numerical domain
limits the elevations to which the fluid elements can sort.  Even without the
effects of a sorting volume with arbitrary height, we should consider the
horizontal extent of the sorting.  By our method, a fluid parcel on the left
side of the domain can sort to a location on the right side of the
domain. \citet{taylor19} compute available potential energy by sorting columns
with horizontal extent of just a few Kolmogorov length scales.  While we point
out that there are other ways to interpret Lorentz's concept of available
potential energy in these simulations, the quantities in figure
\ref{fig:epStrat} are computed consistently for all the cases and we find them
useful for understanding the effects of $\xi$ on energetics, at least
conceptually.


\subsection{Dissipation Rates and Mixing Efficiency}
Mixing is a microscopic process affecting the thermodynamic state of a
fluid. It is irreversible since the fluid cannot be returned to its original,
pre-mixed state without an interaction with the surroundings.  Mixing is
typically quantified by the dissipation rates of kinetic and potential
energy. Here the kinetic energy dissipation rate $\epsilon$ is defined in the
usual way for incompressible flows.

The potential energy dissipation rate, $\epsilon_p$, is the rate at which
available potential energy is irreversibly dissipated to background potential
energy. In some configurations, it might be approximated as proportional to
the diffusive destruction of the variance of $\rho$, which is sometimes
denoted $\chi$.  $\eap$ and $\chi$ are equal only if the reference density
profile (the profile with least potential energy) is equal to the ambient
profile $\overline{\rho}(z)$.  We can write an expression $\eap$ in terms of
$\chi$ plus the difference between $\eap$ and $\chi$ but find it more
informative to follow \citet{scotti14} and write
\begin{equation}
  \eap = -\frac{g}{\rho_0 \Rer\Sc}(|\nabla \rho_t|^2\frac{d z}{d \rho_\ast} -
  |\nabla \rho_\ast|^2\frac{d z}{d \rho_t}) \; .
\label{eq:eap}
\end{equation} 
The first term describes the positive-definite dissipation due to
local irreversible dipycnal mixing \citep[c.f.][]{salehipour16}.  The second term shows that a fluid parcel can
gain or lose available potential energy due to irreversible global
changes in the reference density profile, that is, due to dipycnal mixing
elsewhere in the volume of fluid for which the reference profile is
defined.

Of general interest is how efficiently a flow mixes a scalar field.  One
measure of this is the mixing efficiency
\begin{equation}
\eta = \frac{\eap}{\eap + \epsilon} \, .
\label{eq:eta}
\end{equation}
$\eta$ is of particular interest in field experiments because it is difficult
to measure both $\epsilon$ and $\eap$ simultaneously.  Thus, it is useful to 
measure one quantity (usually $\epsilon$) and relate it
to the other \citep{osborn80}.  \citet{gregg18}
review $\eta$ in the context of ocean measurements and
\citet{taylor19} use DNS data to analyze some of the underlying assumptions in
estimating and applying $\eta$.  Here we compute the volume-averaged mixing
efficiency $\avg{\eta}=\avg{\eap}/\avg{\eap + \epsilon}$ directly from the
simulation data.

\begin{figure}
  \includegraphics{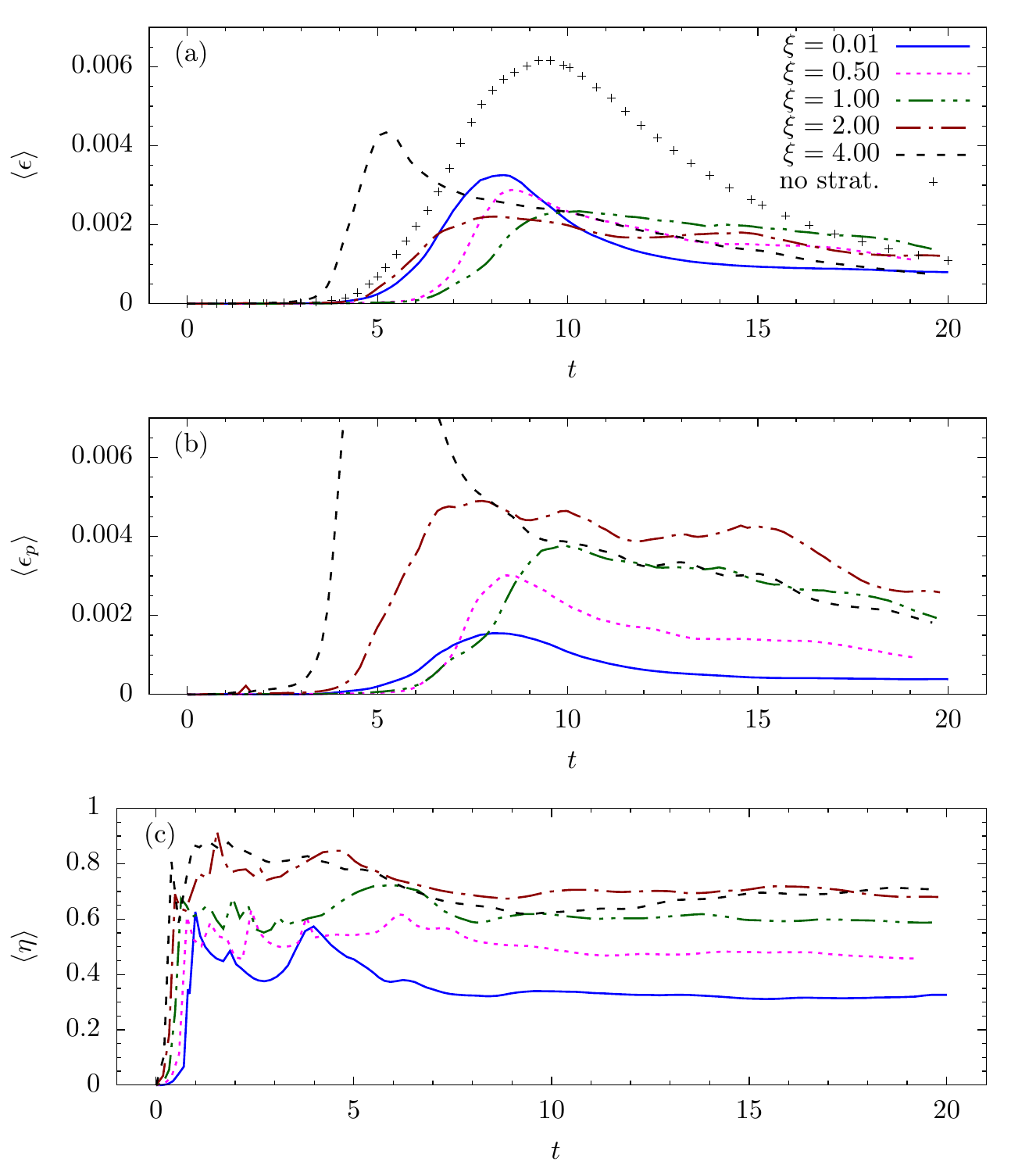}
  \caption{Time evolution of dissipation rates and mixing efficiency.
\label{fig:epsStrat}
\label{fig:chiStrat}}
\end{figure}

The evolutions in time of the boxed-averaged dissipation
rates and mixing efficiency are plotted as figure \ref{fig:epsStrat}.
Recall that the flows are initialised with no small-scale motions and so at
$t=0$ $\avg{\epsilon}$ is very small and $\avg{\eap}$ is zero.  As the flows
evolve, small scale turbulence forms and the dissipation rates increase in
time before decaying.  Any stratification, regardless of $\xi$, reduces
$\avg{\epsilon}$ relative to the unstratified case, which is consistent with
\citet{almalkie12a} and \citet{debk19} for flows that are not viscously
dominated.  There is a weak trend for higher $\xi$ to retard the development
of $\avg{\epsilon}$, except when $\xi=4$, in which case $\avg{\epsilon}$
increases even faster than it does for the unstratified case.
In all the stratified cases, $\avg{\eap}$ follows 
$\avg{\epsilon}$ fairly closely so that $\avg{\eta}$ quickly rises to 
close to its steady state value, and in all cases has settled to the steady
value by about $t=10$.

Considering mixing efficiency further, from figure \ref{fig:epsStrat}(e,f) it
is seen that $\avg{\eta}$ settles by about $t=10$ to much different values when
$\xi < 2$ and $\xi \ge 2$.  For low $\xi$, $\avg{\eta} \approx 0.35$, which is
consistent with simulated flows with $\Sc=1$
\citep{riley03,almalkie12a,debk19}.  The laboratory experiments of
\citet{liu95} indicate a value of 0.41 whereas \citet{maffioli16} find 0.25
for very low Froude number and \citet{osborn80} assumed 0.17 based on ocean
data.  The latter value may be explained by the effects of Schmidt (or
Prandtl) number \citep[c.f.][]{smyth01,salehipour15} or by the effects of forcing from strong vertical mean shear \citep{portwood19}.  

For the simulations with $\xi \ge 2$, $\avg{\eta}\approx 0.65$ when $t>10$.
Two other flows with similarly high $\eta$ is horizontal convection with
$\avg{\eta}=0.85$ \citep{scotti11} and Rayleigh-Taylor and $\avg{\eta}=0.5$
\citep{dalziel08}.  It is not obvious that either of these configurations
relates to ours except, perhaps, that they all generate significant vertical
motion.  Consider again figures \ref{fig:KEStrat} and \ref{fig:evStrat} in
which we observe that vertical motion is induced outside the strongly
stratified region, particularly when $\xi=4$, and results in high $E_a$ when
that motion does work against the strong density gradient near the centreline.
This phenomenon is less pronounced when $\xi=2$, but, nevertheless, apparent.

\section{Modelling Implications and Conclusions}
\label{sec:conclusions}
Simulation results are presented for flows initialised with a von
K\'{a}rm\'{a}n vortex street and no mean velocity or shear, similar to a
momentumless wake. Each simulation is subject to non-uniform density
stratification to represent a wake in natural settings such as that found in
an atmospheric transition layer or thermohaline staircase.  The average
stratification for all the cases is the same so the only simulation parameter
that is adjusted is $\xi$, the ratio of the wake height to the stratification
profile height.  The average stratification is strong enough so that the
average Froude number is order one.  Comparisons to the unstratified case are
also made since it is observed that flows in which the density changes over a
very short vertical range have certain characteristics more like those of
unstratified than stratified flows.  

The simulated flows in which the wake height is less than or equal to twice
the density layer height ($\xi \leq 2$) are observed to be consistent with the
current understanding of strongly stratified flow, in particular,
increasing horizontal and decreasing vertical length scales as the flows
evolve.  These characteristics are in agreement with the scaling arguments of
\cite{riley81} and the horizontal layer decoupling heuristic by
\cite{lilly83}. When the wake height is greater that twice the density layer
height ($\xi > 2$), however, the importance of the density stratification is
diminished and the flows demonstrate characteristics of non-stratified
flows. In particular, horizontal contribution to kinetic energy $\avg{\eh}$
dissipates faster, and the vertical contribution $\avg{\ev}$, as well as the
dissipation rates
$\avg{\epsilon}$ and $\avg{\epsilon_p}$ increase much faster and to much higher
values than in the cases with $\xi \leq 2$.

The transition point of $\xi\approx 2$ suggests that it is the relation
between the stratification profile and the energy profile, rather than the
velocity profile, that determines if the flow will behave primarily as
stratified or unstratified in terms of global statistics.  Rather than simply
concluding that the observed results might be explained by energetics, it
worthwhile to briefly consider instability mechanisms.  The Kelvin-Helmholtz
(KH) instability is typically assumed to be the dominant instability mechanism
in stratified flows. In shear flows where the density length scale becomes
smaller than the velocity length scale, however, theoretical analysis first
performed by \cite{holmboe62} showed an oscillatory instability. Evidence of
the Holmboe instability has been observed in the atmosphere
\citep[e.g.,][]{emmanuel72} and ocean \citep[e.g.,][]{yonemitsu96}. Transition
from KH to Holmboe instability has been investigated theoretically
\citep[e.g.,][]{smyth89, ortiz02}, experimentally
\citep[e.g.,][]{zhu01,hogg03} and numerically
\citep[e.g.,][]{hazel72,smyth88}. In particular, \cite{smyth03} demonstrate
that when the ratio of velocity to density length scales is greater than 2.4
then Holmboe instability becomes dominant.  A significant difference between
those experiments and ours is that our simulations have no mean shear. Also,
we are considering the wake of an object rather than a shear plane.
Nevertheless, the results of \cite{smyth03} support our conclusion that
stratified flow configurations with the equivalent of $\xi > 2$ should not be
expected to behave like uniformly stratified flows.

One motivation for studying flows subject to non-linear stratification is to
improve ocean and atmospheric modelling.  The results of this study suggest
that modelling a sharp localised density gradient with a uniform density
profile having the same average density gradient will result in
underpredicting $\avg{\epsilon_p}$ by a factor of two or more with a
corresponding underprediction of the mixing efficiency.

\begin{acknowledgments} 
  The authors thank Jim Riley and Kraig Winters for the concept of vortex
  street simulations and David Hebert for the preliminary simulations and
  concepts developed with support from Office of Naval Research grant
  N00014-04-1-0687.  This research was supported by Office of Naval Research
  grant N00014-15-1-2248.  High
performance computing resources were provided through the U.S.\ Department of
Defense High Performance Computing Modernization Program by the Army Engineer
Research and Development Center and the Army Research Laboratory under
Frontier Project FP-CFD-FY14-007.  This manuscript is approved for public release by Los Alamos National Laboratory as LA-UR-20-28846.
\end{acknowledgments}

\newpage

\begin{thebibliography}{86}
\expandafter\ifx\csname natexlab\endcsname\relax\def\natexlab#1{#1}\fi

\bibitem[Almalkie \& de~Bruyn~Kops(2012)]{almalkie12a}
{\sc Almalkie, S. \& de~Bruyn~Kops, S.~M.} 2012 Kinetic energy dynamics in
  forced, homogeneous, and axisymmetric stably stratified turbulence. {\em J.
  Turbul.\/} {\bf 13}~(29), 1--29.

\bibitem[Bartello \& Tobias(2013)]{bartello13}
{\sc Bartello, P. \& Tobias, S.~M.} 2013 Sensitivity of stratified turbulence
  to buoyancy {R}eynolds number. {\em J. Fluid Mech.\/} {\bf 725}, 1--22.

\bibitem[Beckers {\em et~al.\/}(2002)Beckers, Clerx, van Heijst \&
  Verzicco]{beckers02}
{\sc Beckers, M., Clerx, H. J.~H., van Heijst, G. J.~F. \& Verzicco, R.} 2002
  Dipole formation by two interacting shielded monopoles in a stratified fluid.
  {\em Phys. Fluids\/} {\bf 14}, 704.

\bibitem[Beckers {\em et~al.\/}(2001)Beckers, Verzicco, Clercx \& van
  Heijst]{beckers01}
{\sc Beckers, M., Verzicco, R., Clercx, H. J.~H. \& van Heijst, G. J.~F.} 2001
  Dynamics of pancake-like vortices in a stratified fluid: experiments, model
  and numerical simulations. {\em J. Fluid Mech.\/} {\bf 433}, 1--27.

\bibitem[Billant \& Chomaz(2000{\natexlab{{\em a\/}}})]{billant00b}
{\sc Billant, P. \& Chomaz, J.-M.} 2000{\natexlab{{\em a\/}}} Experimental
  evidence for a new instability of a vertical columnar vortex pair in a
  strongly stratified fluid. {\em J. Fluid Mech.\/} {\bf 418}, 167--188.

\bibitem[Billant \& Chomaz(2000{\natexlab{{\em b\/}}})]{billant00a}
{\sc Billant, P. \& Chomaz, J.-M.} 2000{\natexlab{{\em b\/}}} Three-dimensional
  stability of a vertical columnar vortex pair in a stratified fluid. {\em J.
  Fluid Mech.\/} {\bf 419}, 65--91.

\bibitem[Billant \& Chomaz(2001)]{billant01}
{\sc Billant, P. \& Chomaz, J.-M.} 2001 Self-similarity of strongly stratified
  inviscid flows. {\em Phys. Fluids\/} {\bf 13}, 1645--1651.

\bibitem[Bonnier \& Eiff(2002)]{bonnier02}
{\sc Bonnier, M. \& Eiff, O.} 2002 Experimental investigation of the collapse
  of a turbulent wake in a stably stratified fluid. {\em Phys. Fluids\/} {\bf
  14}, 791.

\bibitem[Boyd(1989)]{boyd89}
{\sc Boyd, J.~D.} 1989 Properties of thermal staircase off the northeast coast
  of {S}outh {A}merica. {\em J. Geophys. Res.\/} {\bf 94}, 8303--8312.

\bibitem[Boyd(2001)]{boyd01}
{\sc Boyd, J.~P.} 2001 {\em Chebyshev and Fourier Spectral Methods\/}. Dover.

\bibitem[Brethouwer {\em et~al.\/}(2007)Brethouwer, Billant, Lindborg \&
  Chomaz]{brethouwer07}
{\sc Brethouwer, G., Billant, P., Lindborg, E. \& Chomaz, J.-M.} 2007 Scaling
  analysis and simulation of strongly stratified turbulent flows. {\em J. Fluid
  Mech.\/} {\bf 585}, 343--368.

\bibitem[de~Bruyn~Kops(2015)]{debk15}
{\sc de~Bruyn~Kops, S.~M.} 2015 Classical turbulence scaling and intermittency
  in stably stratified {B}oussinesq turbulence. {\em J. Fluid Mech.\/} {\bf
  775}, 436--463.

\bibitem[de~Bruyn~Kops \& Riley(2019)]{debk19}
{\sc de~Bruyn~Kops, S.~M. \& Riley, J.~J.} 2019 The effects of stable
  stratification on the decay of initially isotropic homogeneous turbulence.
  {\em J. Fluid Mech.\/} {\bf 860}, 787–821.

\bibitem[de~Bruyn~Kops {\em et~al.\/}(2003)de~Bruyn~Kops, Riley \&
  Winters]{debk03a}
{\sc de~Bruyn~Kops, S.~M., Riley, J.~J. \& Winters, K.~B.} 2003 {R}eynolds and
  {F}roude number scaling in stably-stratified flows. In {\em {R}eynolds Number
  Scaling in Turbulent Flow\/}. Kluwer.

\bibitem[Chomaz {\em et~al.\/}(1993)Chomaz, Bonneton, Butet \&
  Hopfinger]{cho93b}
{\sc Chomaz, J.-M., Bonneton, P., Butet, A. \& Hopfinger, E.~J.} 1993 Vertical
  diffusion of the far wake of a sphere moving in a stratified fluid. {\em
  Phys. Fluids A\/} {\bf 5}, 2799.

\bibitem[Dalaudier {\em et~al.\/}(1994)Dalaudier, Sidi, Crochet \&
  Vernin]{dalaudier94}
{\sc Dalaudier, F., Sidi, C., Crochet, M. \& Vernin, J.} 1994 Direct evidence
  of "sheets" in the atmospheric temperature field. {\em J. Atmos. Sci.\/} {\bf
  51}, 237--248.

\bibitem[Dalziel {\em et~al.\/}(2008)Dalziel, Patterson, Caulfield \&
  Coomaraswamy]{dalziel08}
{\sc Dalziel, S.~B., Patterson, M.~D., Caulfield, C.~P. \& Coomaraswamy, I.~A.}
  2008 Mixing efficiency in high-aspect-ratio {R}ayleigh-{T}aylor experiments.
  {\em Phys. Fluids\/} {\bf 20}~(6), 065106.

\bibitem[D'Asaro {\em et~al.\/}(2004)D'Asaro, Winters \& Lien]{dasaro04}
{\sc D'Asaro, E.~A., Winters, K.~B. \& Lien, R.~C.} 2004 {L}agrangian estimates
  of diapycnal mixing in a simulated k-h instability. {\em J. Atmos. Oceanic
  Technol.\/} {\bf 21}, 799--809.

\bibitem[Diamessis {\em et~al.\/}(2011)Diamessis, Spedding \&
  Domaradzki]{diamessis11}
{\sc Diamessis, P.~J., Spedding, G.~R. \& Domaradzki, J.~A.} 2011 Similarity
  scaling and vorticity structure in high-{R}eynolds-number stably stratified
  turbulent wakes. {\em J. Fluid Mech.\/} {\bf 671}, 52--95.

\bibitem[Dillon \& Caldwell(1980)]{dillon80}
{\sc Dillon, T.~M. \& Caldwell, D.~R.} 1980 The {B}atchelor spectrum and
  dissipation in the upper ocean. {\em J. Geophysical Research\/} {\bf
  85}~(C4), 1910--1916.

\bibitem[Emmanuel {\em et~al.\/}(1972)Emmanuel, Bean, McAllister \&
  Pollard]{emmanuel72}
{\sc Emmanuel, C.~B., Bean, B., McAllister, R. \& Pollard, J.~R.} 1972
  Observations of helmholtz waves in the lower atmosphere with an accoustic
  sounder. {\em J. Atmos. Sci.\/} {\bf 29}, 886--892.

\bibitem[Eswaran \& Pope(1988)]{eswaran88}
{\sc Eswaran, V. \& Pope, S.~B.} 1988 Direct numerical simulations of the
  turbulent mixing of a passive scalar. {\em Phys. Fluids\/} {\bf 31},
  506--520.

\bibitem[Falder {\em et~al.\/}({2016})Falder, White \& Caulfield]{falder16}
{\sc Falder, M., White, N.~J. \& Caulfield, C.~P.} {2016} {Seismic imaging of
  rapid onset of stratified turbulence in the south {A}tlantic {O}cean}. {\em
  J. Phys. Oceanogr.\/} {\bf {46}}~({4}), {1023--1044}.

\bibitem[Fincham {\em et~al.\/}(1996)Fincham, Maxworthy \& Spedding]{fincham96}
{\sc Fincham, A.~M., Maxworthy, T. \& Spedding, G.~R.} 1996 Energy dissipation
  and vortex structure in freely decaying, stratified grid turbulence. {\em
  Dyn. Atmos. Oceans\/} {\bf 23}, 155--169.

\bibitem[Gargett {\em et~al.\/}(1984)Gargett, Osborn \& Nasmyth]{gargett84}
{\sc Gargett, A., Osborn, T. \& Nasmyth, P.} 1984 Local isotropy and the decay
  of turbulence in a stratified fluid. {\em J. Fluid Mech.\/} {\bf 144},
  231--280.

\bibitem[Gargett {\em et~al.\/}(2003)Gargett, Merryfield \&
  Holloway]{gargett03}
{\sc Gargett, A.~E., Merryfield, W.~J. \& Holloway, G.} 2003 Direct numerical
  simulation of differential scalar diffusion in three-dimensional stratified
  turbulence. {\em J. Phys. Oceanogr.\/} {\bf 33}, 1758--1782.

\bibitem[Gibson(1980)]{gibson80}
{\sc Gibson, C.~H.} 1980 Fossil turbulence, salinity, and vorticity turbulence
  in the ocean. In {\em Marine Turbulence\/} (ed. J.~C. Nihous), pp. 221--257.
  Elsevier.

\bibitem[Gossard {\em et~al.\/}(1985)Gossard, Gaynor, Zamora \&
  Neff]{gossard85}
{\sc Gossard, E.~E., Gaynor, J.~E., Zamora, R.~J. \& Neff, W.~D.} 1985
  Finestructure of elevated stable layers observed by sounder and in situ tower
  sensors. {\em J. Atmos. Sci.\/} {\bf 4}, 113--131.

\bibitem[Gregg {\em et~al.\/}(2018)Gregg, {D'}Asaro,  \& Riley]{gregg18}
{\sc Gregg, M.~C., {D'}Asaro, E.,  \& Riley, J.} 2018 Mixing coefficients and
  mixing efficiency in the ocean. {\em Annu. Rev. Marine Sci.\/} {\bf 10},
  443--473.

\bibitem[Gregg \& Sanford(1987)]{gregg87a}
{\sc Gregg, M.~C. \& Sanford, T.} 1987 Shear and turbulence in a thermohaline
  staircase. {\em Deep-Sea Research\/} {\bf 34}, 1689--1696.

\bibitem[Hazel(1972)]{hazel72}
{\sc Hazel, P.} 1972 Numerical studies of the stability of inviscid parallel
  shear flows. {\em J. Fluid Mech.\/} {\bf 51}, 39--62.

\bibitem[Hebert(2007)]{hebert07}
{\sc Hebert, D.~A.} 2007 Mixing in stably stratified flows. PhD thesis,
  University of Massachusetts Amherst.

\bibitem[Hebert \& de~Bruyn~Kops(2006)]{hebert06a}
{\sc Hebert, D.~A. \& de~Bruyn~Kops, S.~M.} 2006 Relationship between vertical
  shear rate and kinetic energy dissipation rate in stably stratified flows.
  {\em Geophys. Res. Let.\/} {\bf 33}, L06602.

\bibitem[Herring \& M\'etais(1989)]{herring89}
{\sc Herring, J.~R. \& M\'etais, O.} 1989 Numerical experiments in forced
  stably stratified turbulence. {\em J. Fluid Mech.\/} {\bf 202}, 97--115.

\bibitem[Hogg \& Ivey(2003)]{hogg03}
{\sc Hogg, A.~M. \& Ivey, G.~N.} 2003 The kelvin-helmholtz to holmboe
  instability transition in stratified exchange flows. {\em J. Fluid Mech.\/}
  {\bf 477}, 339--362.

\bibitem[Holliday \& McIntyre(1981)]{holliday81}
{\sc Holliday, D. \& McIntyre, M.} 1981 On potential energy density in an
  incompressible, stratified fluid. {\em J. Fluid Mech.\/} {\bf 107}, 221--225.

\bibitem[Holmboe(1962)]{holmboe62}
{\sc Holmboe, J.} 1962 On the behavior of symmetric waves in stratified shear
  layers. {\em Geophys. Publ.\/} {\bf 24}, 67--113.

\bibitem[Ivey \& Imberger(1991)]{ivey91}
{\sc Ivey, G.~N. \& Imberger, J.} 1991 On the nature of turbulence in a
  stratified fluid. part 1: The energetics of mixing. {\em J. Phys.
  Oceanogr.\/} {\bf 21}, 650--658.

\bibitem[Jang \& de~Bruyn~Kops(2007)]{jang07}
{\sc Jang, Y. \& de~Bruyn~Kops, S.~M.} 2007 Pseudo-spectral numerical
  simulation of miscible fluids with a high density ratio. {\em Comput.
  Fluids\/} {\bf 36}, 238--247.

\bibitem[Kaneda {\em et~al.\/}({2003})Kaneda, Ishihara, Yokokawa, Itakura \&
  Uno]{kaneda03}
{\sc Kaneda, Y., Ishihara, T., Yokokawa, M., Itakura, K. \& Uno, A.} {2003}
  {Energy dissipation rate and energy spectrum in high resolution direct
  numerical simulations of turbulence in a periodic box}. {\em Phys. Fluids\/}
  {\bf {15}}~({2}), {L21--L24}.

\bibitem[Kerr(1985)]{ker85}
{\sc Kerr, R.~M.} 1985 Higher-order derivative correlations and the alignment
  of small-scale structures in isotropic turbulence. {\em J. Fluid Mech.\/}
  {\bf 153}~(31), 31.

\bibitem[Lambert \& Sturges(1977)]{lambert77}
{\sc Lambert, R.~B. \& Sturges, W.} 1977 A thermohaline staircase and vertical
  mixing in the thermocline. {\em Deep-Sea Research\/} {\bf 24}, 211--222.

\bibitem[Lilly(1983)]{lilly83}
{\sc Lilly, D.~K.} 1983 Stratified turbulence and the mesoscale variability of
  the atmosphere. {\em J. Atmos. Sci.\/} {\bf 40}, 749--761.

\bibitem[Lindborg(2006)]{lindborg06a}
{\sc Lindborg, E.} 2006 The energy cascade in a strongly stratified fluid. {\em
  J. Fluid Mech.\/} {\bf 550}, 207--242.

\bibitem[Liu(1995)]{liu95}
{\sc Liu, H.~T.} 1995 Energetics of grid turbulence in a stably stratified
  fluid. {\em J. Fluid Mech.\/} {\bf 296}, 127--157.

\bibitem[Lorenz(1955)]{lorenz55}
{\sc Lorenz, E.~N.} 1955 Available potential energy and the maintenance of the
  general circulation. {\em Tellus\/} {\bf 7}, 157--167.

\bibitem[Maffioli \& Davidson(2016)]{maffioli16}
{\sc Maffioli, A. \& Davidson, P.~A.} 2016 Dynamics of stratified turbulence
  decaying from a high buoyancy {R}eynolds number. {\em J. Fluid Mech.\/} {\bf
  786}, 210--233.

\bibitem[M\'etais \& Herring(1989)]{metais89}
{\sc M\'etais, O. \& Herring, J.~R.} 1989 Numerical simulations of freely
  evolving turbulence in stably stratified fluids. {\em J. Fluid Mech.\/} {\bf
  202}, 117--148.

\bibitem[Meunier \& Spedding(2006)]{meunier06a}
{\sc Meunier, P. \& Spedding, G.} 2006 Stratified propelled wakes. {\em J.
  Fluid Mech.\/} {\bf 522}, 229--256.

\bibitem[Molcard \& Tait(1977)]{molcard77}
{\sc Molcard, R. \& Tait, R.~I.} 1977 The steady state of the step structure in
  the {T}yrrhenian {S}ea. In {\em A voyage of discovery\/} (ed. M.~V. Angel),
  pp. 221--233. New York:Pergamon Press.

\bibitem[Molemaker \& McWilliams(2010)]{molemaker10}
{\sc Molemaker, M.~J. \& McWilliams, J.~C.} 2010 Local balance and cross-scale
  flux of available potential energy. {\em J. Fluid Mech.\/} {\bf 645},
  295--314.

\bibitem[Muschinski \& Wode(1998)]{muschinski98}
{\sc Muschinski, A. \& Wode, C.} 1998 First in situ evidence for coexisting
  submeter temperature and humidity sheets in the lower free troposphere. {\em
  J. Atmos. Sci.\/} {\bf 55}~(18), 2893--2905.

\bibitem[Ortiz {\em et~al.\/}(2002)Ortiz, Chomaz \& Loiseleux]{ortiz02}
{\sc Ortiz, S., Chomaz, J.-M. \& Loiseleux, T.} 2002 Spatial holmboe
  instability. {\em Phys. Fluids\/} {\bf 14}~(8), 2585--2597.

\bibitem[Osborn(1980)]{osborn80}
{\sc Osborn, T.~R.} 1980 Estimates of the local-rate of vertical diffusion from
  dissipation measurements. {\em J. Phys. Oceanogr.\/} {\bf 10}, 83--89.

\bibitem[Peltier \& Caulfield(2003)]{peltier03}
{\sc Peltier, W.~R. \& Caulfield, C.~P.} 2003 Mixing efficiency in stratified
  shear flows. {\em Annu. Rev. Fluid Mech.\/} {\bf 35}, 135--167.

\bibitem[Portwood {\em et~al.\/}(2019)Portwood, de~Bruyn~Kops \&
  Caulfield]{portwood19}
{\sc Portwood, G., de~Bruyn~Kops, S. \& Caulfield, C.} 2019 Asymptotic dynamics
  of high dynamic range stratified turbulence. {\em Phys. Rev. Lett.\/} {\bf
  122}~(19), 194504.

\bibitem[Portwood {\em et~al.\/}(2016)Portwood, de~Bruyn~Kops, Taylor,
  Salehipour \& Caulfield]{portwood16}
{\sc Portwood, G.~D., de~Bruyn~Kops, S.~M., Taylor, J.~R., Salehipour, H. \&
  Caulfield, C.~P.} 2016 Robust identification of dynamically distinct regions
  in stratified turbulence. {\em J. Fluid Mech.\/} {\bf 807}, R2 (14 pages).

\bibitem[Praud {\em et~al.\/}(2005)Praud, Fincham \& Sommeria]{praud05}
{\sc Praud, O., Fincham, A.~M. \& Sommeria, J.} 2005 Decaying grid turbulence
  in a strongly stratified fluid. {\em J. Fluid Mech.\/} {\bf 522}, 1--33.

\bibitem[Riley \& de~Bruyn~Kops(2003)]{riley03}
{\sc Riley, J.~J. \& de~Bruyn~Kops, S.~M.} 2003 Dynamics of turbulence strongly
  influenced by buoyancy. {\em Phys. Fluids\/} {\bf 15}~(7), 2047--2059.

\bibitem[Riley \& Lelong(2000)]{riley00}
{\sc Riley, J.~J. \& Lelong, M.~P.} 2000 Fluid motions in the presence of
  strong stable stratification. {\em Annu. Rev. Fluid Mech.\/} {\bf 32},
  613--657.

\bibitem[Riley {\em et~al.\/}(1981)Riley, Metcalfe \& Weissman]{riley81}
{\sc Riley, J.~J., Metcalfe, R.~W. \& Weissman, M.~A.} 1981 Direct numerical
  simulations of homogeneous turbulence in density stratified flows. In {\em
  Proc. AIP Conf. Nonlinear Properties of Internal Waves\/} (ed. B.~J. West),
  pp. 79--112. New York: American Institute of Physics.

\bibitem[Roullet \& Klein(2009)]{roullet09}
{\sc Roullet, G. \& Klein, P.} 2009 Available potential energy diagnosis in a
  direct numerical simulation of rotating stratified turbulence. {\em J. Fluid
  Mech.\/} {\bf 624}, 45--55.

\bibitem[Salehipour \& Peltier(2015)]{salehipour15}
{\sc Salehipour, H. \& Peltier, W.} 2015 Diapycnal diffusivity, turbulent
  {P}randtl number and mixing efficiency in {B}oussinesq stratified turbulence.
  {\em J. Fluid Mech.\/} {\bf 775}, 464--500.

\bibitem[Salehipour {\em et~al.\/}(2016)Salehipour, Peltier, Whalen \&
  MacKinnon]{salehipour16}
{\sc Salehipour, H., Peltier, W.~R., Whalen, C.~B. \& MacKinnon, J.~A.} 2016 A
  new characterization of the turbulent diapycnal diffusivities of mass and
  momentum in the ocean. {\em Geophys. Res. Lett.\/} {\bf 43}~(7), 3370--3379.

\bibitem[Schmitt {\em et~al.\/}(1987)Schmitt, Perkins, Boyd \&
  Stalcup]{schmitt87}
{\sc Schmitt, R.~W., Perkins, H., Boyd, J.~D. \& Stalcup, M.~C.} 1987 {C-SALT}:
  An investigation of the thermohaline staircase in the western tropical
  {N}orth {A}tlantic. {\em Deep-Sea Research\/} {\bf 34}, 1655--1665.

\bibitem[Scotti \& White(2011)]{scotti11}
{\sc Scotti, A. \& White, B.} 2011 Is horizontal convection really
  “non-turbulent?”. {\em Geophys. Res. Lett.\/} {\bf 38}~(21).

\bibitem[Scotti \& White(2014)]{scotti14}
{\sc Scotti, A. \& White, B.} 2014 Diagnosing mixing in stratified turbulent
  flows with a locally defined available potential energy. {\em J. Fluid
  Mech.\/} {\bf 740}, 114.

\bibitem[Shih {\em et~al.\/}(2005)Shih, Koseff, Ivey \& Ferziger]{shih05}
{\sc Shih, L.~H., Koseff, J.~R., Ivey, G.~N. \& Ferziger, J.~H.} 2005
  Parameterization of turbulent fluxes and scales using homogeneous sheared
  stably stratified turbulence simulations. {\em J. Fluid Mech.\/} {\bf 525},
  193--214.

\bibitem[Smyth {\em et~al.\/}(1988)Smyth, Klaassen \& Peltier]{smyth88}
{\sc Smyth, W.~D., Klaassen, G.~P. \& Peltier, W.~R.} 1988 Finite-amplitude
  holmboe waves. {\em Geophys. Astrophys. Fluid Dyn.\/} {\bf 43}, 181--222.

\bibitem[Smyth \& Moum(2000{\natexlab{{\em a\/}}})]{smyth00b}
{\sc Smyth, W.~D. \& Moum, J.~N.} 2000{\natexlab{{\em a\/}}} Anisotropy of
  turbulence in stably stratified mixing layers. {\em Phys. Fluids\/} {\bf 12},
  1343--1362.

\bibitem[Smyth \& Moum(2000{\natexlab{{\em b\/}}})]{smyth00a}
{\sc Smyth, W.~D. \& Moum, J.~N.} 2000{\natexlab{{\em b\/}}} Length scales of
  turbulence in stably stratified mixing layers. {\em Phys. Fluids\/} {\bf 12},
  1327--1342.

\bibitem[Smyth {\em et~al.\/}(2001)Smyth, Moum \& Caldwell]{smyth01}
{\sc Smyth, W.~D., Moum, J.~N. \& Caldwell, D.~R.} 2001 The efficiency of
  mixing in turbulent patches: inferences from direct simulations and
  microstructure observations. {\em J. Phys. Oceanogr.\/} {\bf 31}, 1969--1992.

\bibitem[Smyth {\em et~al.\/}(2005)Smyth, Nash \& Moum]{smyth05}
{\sc Smyth, W.~D., Nash, J.~D. \& Moum, J.~N.} 2005 Differential diffusion in
  breaking kelvin-helmholtz billows. {\em J. Phys. Oceanogr.\/} {\bf 35},
  1004--1022.

\bibitem[Smyth \& Peltier(1989)]{smyth89}
{\sc Smyth, W.~D. \& Peltier, W.~R.} 1989 The transition between
  kelvin-helmholtz and holmboe instability - an investigation of the
  overreflection hypothesis. {\em J. Atmos. Sci.\/} {\bf 46}, 3698--3720.

\bibitem[Smyth \& Winters(2003)]{smyth03}
{\sc Smyth, W.~D. \& Winters, K.~B.} 2003 Turbulence and mixing in holmboe
  waves. {\em J. Phys. Oceanogr.\/} {\bf 33}, 694--711.

\bibitem[Spedding(2002)]{spedding02}
{\sc Spedding, G.~R.} 2002 Vertical structure in stratified wakes at high
  initial {F}roude number. {\em J. Fluid Mech.\/} {\bf 454}, 71.

\bibitem[Spedding {\em et~al.\/}(1996)Spedding, Browand \&
  Fincham]{spedding96b}
{\sc Spedding, G.~R., Browand, F.~K. \& Fincham, A.~M.} 1996 Turbulence,
  similarity scaling and vortex geometry in the wake of a towed sphere in a
  stably stratified fluid. {\em J. Fluid Mech.\/} {\bf 314}, 53.

\bibitem[Staquet(2000)]{staquet00}
{\sc Staquet, C.} 2000 Mixing in a stably-stratified shear layer: two- and
  three-dimensional numerical experiments. {\em Fl. Dyn. Res.\/} {\bf 27}, 367.

\bibitem[Taylor {\em et~al.\/}(2019)Taylor, de~Bruyn~Kops, Caulfield \&
  Linden]{taylor19}
{\sc Taylor, J.~R., de~Bruyn~Kops, S.~M., Caulfield, C.~P. \& Linden, P.~F.}
  2019 Testing the assumptions underlying ocean mixing methodologies using
  direct numerical simulations. {\em J. Phys. Oceanogr.\/} {\bf 49}~(11),
  2761--2779.

\bibitem[Waite \& Bartello(2004)]{waite04}
{\sc Waite, M. \& Bartello, P.} 2004 Stratified turbulence dominated by
  vortical motion. {\em J. Fluid Mech.\/} {\bf 517}, 281--308.

\bibitem[Watanabe {\em et~al.\/}(2016)Watanabe, Riley, de~Bruyn~Kops, Diamessis
  \& Zhou]{watanabe16}
{\sc Watanabe, T., Riley, J.~J., de~Bruyn~Kops, S.~M., Diamessis, P.~J. \&
  Zhou, Q.} 2016 Turbulent/non-turbulent interfaces in wakes in stably
  stratified fluids. {\em J. Fluid Mech.\/} {\bf 797}, R1.

\bibitem[Williams(1974)]{williams74}
{\sc Williams, A.~J.} 1974 Salt fingers observed in the {M}editerranean
  outflow. {\em Science\/} {\bf 185}, 941--943.

\bibitem[Winters \& Barkan(2013)]{winters13}
{\sc Winters, K.~B. \& Barkan, R.} 2013 Available potential energy density for
  boussinesq fluid flow. {\em J. Fluid Mech.\/} {\bf 714}, 476--488.

\bibitem[Winters \& D'Asaro(1996)]{winters96}
{\sc Winters, K.~B. \& D'Asaro, E.~A.} 1996 Diascalar flux and the rate of
  fluid mixing. {\em J. Fluid Mech.\/} {\bf 317}, 179--193.

\bibitem[Yonemitsu {\em et~al.\/}(1996)Yonemitsu, Swaters, Rajaratnam \&
  Lawrence]{yonemitsu96}
{\sc Yonemitsu, N., Swaters, G., Rajaratnam, N. \& Lawrence, G.} 1996 Shear
  instabilities in arrested salt-wedge flows. {\em Dyn. Atmos. Oceans\/} {\bf
  24}, 173--182.

\bibitem[Zhu \& Lawrence(2001)]{zhu01}
{\sc Zhu, D. \& Lawrence, G.} 2001 Holmboe's instability in exchange flows.
  {\em J. Fluid Mech.\/} {\bf 429}, 391--409.

\end{thebibliography}
\clearpage

\end{document}